\documentclass[11pt,draftclsnofoot,journal,twoside,onecolumn,romanappendices]{IEEEtran}
\IEEEoverridecommandlockouts
\usepackage[dvips]{psfrag}
\usepackage[dvips]{epsfig}
\usepackage{epic,color}
\usepackage[cmex10]{amsmath}
\usepackage{amsfonts}
\usepackage{amssymb}
\usepackage{amsmath}
\usepackage{yfonts}
\usepackage{array}
\usepackage{bigstrut}
\usepackage{multirow}
\usepackage{dsfont}
\usepackage{cite}
\usepackage{fixltx2e}
\usepackage{accents}
\usepackage[normalem]{ulem}
\usepackage{enumerate}
\usepackage{stfloats}
\usepackage[T1]{fontenc}
\usepackage{pgf,tikz}
\usepackage{float}
\usepackage{epstopdf}
\usetikzlibrary{decorations.markings,arrows}
\usetikzlibrary{arrows.meta}
\usetikzlibrary{shapes.geometric}
\usepackage{acronym}
\usepackage{flushend}
\usepackage{mathtools}
\usepackage{booktabs,subcaption,amsfonts,dcolumn}
\newcolumntype{d}[1]{D..{#1}}

\DeclarePairedDelimiter\ceil{\lceil}{\rceil}

\newtheorem{definition}{Definition}

\acrodef{WSN}{wireless sensor network}
\acrodef{HARQ}{hybrid automatic repeat request}
\acrodef{ARQ}{automatic repeat request}
\acrodef{FEC}{forward error correction}
\acrodef{SS}{selective sampling}
\acrodef{SSI}{SS information}
\acrodef{CSI}{channel state information}
\acrodef{BSI}{battery state information}
\acrodef{EHN}{energy harvesting node}
\acrodef{PDP}{packet drop probability}
\acrodef{PEP}{packet error probability}
\acrodef{BEP}{bit error probability}
\acrodef{MDP}{Markov decision process}
\acrodef{POMDP}{partially observable Markov decision process}
\acrodef{MLPH}{maximum-likelihood policy heuristic}
\acrodef{ACK}{acknowledgement}
\acrodef{AWGN}{additive white Gaussian noise}
\acrodef{NAK}{negative acknowledgement}
\acrodef{PDF}{probability density function}
\acrodef{BPSK}{binary phase ship keying}
\acrodef{QAM}{quadrature amplitude modulation}
\acrodef{i.i.d.}{independent and identically distributed}
\acrodef{FSMC}{finite state Markov chain}
\acrodef{EH}{energy harvesting}
\acrodef{SNR}{signal-to-noise ratio}

\begin{document}

\title{\noindent Energy Management for Energy Harvesting Wireless Sensors with Adaptive Retransmission}

\author{\noindent \begingroup\centering{Animesh Yadav,~\IEEEmembership{Member,~IEEE}, Mathew Goonewardena,~\IEEEmembership{Student Member, IEEE}, Wessam Ajib, \IEEEmembership{Senior Member, IEEE}, Octavia A. Dobre,~\IEEEmembership{Senior Member, IEEE}, and Halima Elbiaze,~\IEEEmembership{Member, IEEE}}\endgroup%
\thanks{A part of the paper is published in the Proceeding of the IEEE International Conference on Communications (IEEE ICC 2015), London, UK, 8-12 June 2015.

A. Yadav and O. A. Dobre are with the Faculty of Engineering and Applied Science, Memorial University, St. John's, NL, Canada, (e-mail: \{animeshy, odobre\}@mun.ca), W. Ajib and H. Elbiaze are with the Department of Computer Science, Universit\'e du Qu\'ebec \`a Montr\'eal (UQAM), Montreal, QC, Canada (e-mail: \{ajib.wessam, elbiaze.halima\}@uqam.ca) and M. Goonewardena is \'Ecole de Technologie Sup\'erieure (\'ETS), Montr\'eal, QC, Canada. (e-mail:mathew-pradeep.goonewardena.1@ens.etsmtl.ca).%
}}
\maketitle
\vspace{-1in}
\begin{abstract}
This paper analyzes the communication between two energy harvesting wireless sensor nodes. The nodes use automatic repeat request and forward error correction mechanism for the error control. \textcolor{black}{The random nature of available energy and arrivals of harvested energy may induces interruption to the signal sampling and decoding operations. We} propose a selective sampling scheme where the length of the transmitted packet to be sampled depends on the available energy at the receiver. The receiver performs the decoding when complete samples of the packet are available. The selective sampling information bits are piggybacked on the automatic repeat request messages for the transmitter use. This way, the receiver node \textcolor{black}{manages more efficiently its energy use.} Besides, we present the partially observable Markov decision process formulation, which minimizes the long-term average pairwise error probability and optimizes the transmit power. \textcolor{black}{Optimal} and suboptimal power assignment strategies \textcolor{black}{are introduced for retransmissions, which are adapted to the selective sampling and channel state information.} With finite battery size and fixed power assignment policy, an analytical expression for the average \acl{PDP} is derived. Numerical simulations show the performance gain of the proposed scheme with power assignment strategy over the conventional scheme.
\end{abstract}

\IEEEpeerreviewmaketitle{\noindent }

\section{Introduction}
The use of \ac{EH} sources to power wireless communication systems has recently received considerable attention \cite{Ozel-Tutuncuoglu-Yang-Ulukus-Yener-jsac-11,Gorlatova-Wallwater-Zussman-infocom-11,Yang-Ulukus-tcom-12, Tutuncuoglu-Yener-ita-12,Mahdavi-Doost-Yates-isit-13,Yates-Mahdavi-Doost-gsip-13,Mahdavi-Doost-Yates-ciss-14}. The EH devices offer green communication and can operate autonomously over long periods of time. Because of these benefits, the EH devices are
also increasingly considered in \acp{WSN} to power the sensor nodes \cite{Kansal-Hsu-Zahedi-Srivastava-TECS-07,Murthy-ijwin-09, Shenqiu-Seyedi-Sikdar-twcom-13,Aprem-Murthy-Mehta-jstsp-13, Zhou-Chen-Chen-Niu-jsac-15, Sharma-Murthy-gsip-14,Doshi-Vaze-iccs-14,Yadav-Goonewardhena-Ajib-Elbiaze-icc-15}.Sensors nodes are low cost distributed \textcolor{black}{devices which} operate on minimal energy. They are very prevalent in applications related to monitoring and controlling the environments, especially the remote and dangerous ones \cite{Stankovic-Abdelzaher-Chenyang-Lui-Hou-proc-03}. Usually, sensor nodes are operated by small capacity non-renewable batteries, thus, suffering from finite lifespan of operation. Sensor nodes with \ac{EH} capabilities can be an alternative to increase the lifespan and lower the maintenance cost. Energy can be harvested from the environment using for instance solar, vibration or thermoelectric effects. \textcolor{black}{Unlike \ac{EH}, another practical alternative to increase the lifespan of the nodes is to use massive antenna arrays at the receiver nodes to mitigates severe energy constraints given by the inexpensive transmitter nodes \cite{Ciuonzo-Rossi-Dey-tsp-15, Shirazinia-Dey-Ciuonzo-Rossi-tsp-16}.}

Typically, \textcolor{black}{the} energy arrival amount at the EH devices is random. Thus, for such nodes, the challenging objective is the adequate \textcolor{black}{management} of the collected energy to enable reliable and continuous operation. \textcolor{black}{Recently, a} considerable amount of works on wireless networks solely powered by harvested energy have emanated \cite{Kansal-Hsu-Zahedi-Srivastava-TECS-07,Murthy-ijwin-09,Aprem-Murthy-Mehta-jstsp-13,Shenqiu-Seyedi-Sikdar-twcom-13,Mahdavi-Doost-Yates-isit-13,Yates-Mahdavi-Doost-gsip-13,Mahdavi-Doost-Yates-ciss-14,Tutuncuoglu-Yener-ita-12} to address this objective. Although both transmitter and receiver nodes can harvest energy, the research is primarily focused either on the transmitter \cite{Aprem-Murthy-Mehta-jstsp-13,Kansal-Hsu-Zahedi-Srivastava-TECS-07,Murthy-ijwin-09,Shenqiu-Seyedi-Sikdar-twcom-13} or receiver \cite{Mahdavi-Doost-Yates-isit-13,Yates-Mahdavi-Doost-gsip-13,Bai-Mezghani-Nossek-wsa-13,Mahdavi-Doost-Yates-ciss-14}. \textcolor{black}{There are many practicals scenarios where the transmitter and receiver nodes can harvest energy to increase their lifespan, such as the scenario of transmitter and multiple intermediate nodes in a multi-hop \ac{WSN}, and multiple transmitter nodes communicating with a single sink node. These scenarios are more challenging due to the presence of many random sources of energy.}

Fewer works have considered the \ac{EH} capability at both transmitter and receiver nodes simultaneously \cite{Tutuncuoglu-Yener-ita-12,Zhou-Chen-Chen-Niu-jsac-15,Sharma-Murthy-gsip-14,Doshi-Vaze-iccs-14,Yadav-Goonewardhena-Ajib-Elbiaze-icc-15}. In \cite{Tutuncuoglu-Yener-ita-12}, the authors considered a static \ac{AWGN} channel and used a rate-based utility as a function of both transmitter and receiver powers. They proposed directional water-filling based power allocation policy in an offline setting.  The problem of online power control for a wireless link with \ac{ARQ} scheme is studied in \cite{Zhou-Chen-Chen-Niu-jsac-15}. The authors investigated three fixed policies under various assumptions, such as knowledge of the receiver battery availability at the transmitter node, finite and infinite battery storage. In \cite{Sharma-Murthy-gsip-14}, \textcolor{black}{the} authors analyzed a wireless link which \textcolor{black}{employs} type-II \ac{HARQ} scheme. They derived the \ac{PDP} for predetermined transmit energy levels. In \cite{Doshi-Vaze-iccs-14}, the authors \textcolor{black}{obtained} a lower bound on maximum achievable throughput and proposed a common threshold policy. In \cite{Yadav-Goonewardhena-Ajib-Elbiaze-icc-15}, we introduced \textcolor{black}{an} energy-aware adaptive retransmission scheme, where the receiver node performs the \ac{SS} and decoding operations based on the energy availability. 

The receiver node \textcolor{black}{spends} the energy dominantly in sampling and decoding operations, if a \ac{FEC} coding is employed \cite{Cui-Goldsmith-Bahai-jsac-04}. Moreover, for small distances, \textcolor{black}{the} transmit energy is often smaller than the energy needed in the decoding operation \cite{Grover-Woyach-Sahai-jsac-11}. Nonetheless, \textcolor{black}{because of the randomness} in the amount of energy arrivals, the receiver operations \textcolor{black}{may be suspended, which} leads to energy wastage. Thus, the receiver might favor to sample a fraction of the full packet depending on the available energy \cite{Mahdavi-Doost-Yates-isit-13}, which we \textcolor{black}{refer to as} \ac{SS}. On the other hand, the transmitter with exact \ac{SSI} can retransmit only a portion of the packet, which is not sampled by the receiver. 

Furthermore, the \textcolor{black}{time-varying} characteristic of the wireless channel and harvested energy might contribute to a higher \ac{PEP}. Hence, the transmitter must adequately \textcolor{black}{adapt} the transmit power level to the \ac{CSI}, while meeting the  constraint of energy causality, to \textcolor{black}{ensure} a lower \ac{PEP}. \textcolor{black}{The} causality constraint affirms that the cumulative \textcolor{black}{used} energy cannot surpass the cumulative harvested energy by nodes at any given time. Furthermore, based on the \ac{SSI} knowledge, the transmitter adapts the packet size to ensure an efficient utilization of the receiver energy. In pursuance of providing the \ac{SSI} to the transmitter, we resort to the \ac{ARQ} protocol's \ac{ACK}/\ac{NAK} feedback messages. Consequently, \textcolor{black}{the retransmission scheme, which we denote by \ac{ACK}/NAKx, needs to have some additional feedback messages}.

In this paper, we consider \textcolor{black}{a generic} communication between two \ac{EH} wireless nodes with \textcolor{black}{the} aforementioned retransmission protocol.  A decision-theoretic approach is used to find the optimal transmit power strategy. Firstly, the problem is formulated as a \ac{POMDP}, which is \textcolor{black}{a} suitable approach for formulating problems that require sequential decision making in a stochastic setting, when some of the system states are unknown \cite{Putterman-94}. We solve the \ac{POMDP} problem using the value iteration method by computing the value function for the belief of the unknown state. Since the memory and computational complexity requirements are \textcolor{black}{limited} for sensor nodes, we propose a suboptimal and a computationally lower greedy power assignment method. 

The outline of this paper is as follows. The system model is presented in Section II. The adaptive retransmission
scheme is detailed in Section III. Section IV introduces \textcolor{black}{the} optimal and suboptimal methods aiming to allocate the power over the time slots. An analytical upper bound on the \ac{PDP} is derived in Section V. Simulated numerical results and discussions are presented in Section VI, followed by conclusions in Section VII. \textcolor{black}{A list of symbols with their descriptions used in this paper is given in Table~\ref{tab:symbolslist} }

\begin{figure}[h!]
\begin{tikzpicture}
\scriptsize
\draw (0,0) rectangle (6,1);
\draw (1,0) -- (1,1);
\draw (2,0) -- (2,1);
\draw (3,0) -- (3,1);
\draw (4,1) -- (4,0);
\draw (5,0) -- (5,1);
\draw (6,0) -- (6,1);
\draw[loosely dotted] (7.0,.5) -- (7.5,.5);
\draw[->] (0,0) -- (8.5,0);
\draw [draw=black, fill=yellow, opacity=0.2] (0,0) rectangle (1,1);  
\draw [draw=black, fill=yellow, opacity=0.2] (1,0) rectangle (1.75,1);  
\draw [draw=black, fill=yellow, opacity=0.2] (2,0) rectangle (2.25,1);   
\draw [draw=black, fill=yellow, opacity=0.2] (3,0) rectangle (4,1);

\draw[<-] (0,1) -- (0,2) node [midway,fill=white] {$E_{\text{Tx}}^{\text{h}}$};
\draw[<-,dashed] (0,2) -- (0,3) node [midway,fill=white] {ACK};
\draw[<-] (1,1) -- (1,2) node [midway,fill=white] {$E_{\text{Tx}}^{\text{h}}$};
\draw[<-,dashed] (1,2) -- (1,3) node [midway,fill=white] {NAK};
\draw[<-] (2,1) -- (2,2) node [midway,fill=white] {X};
\draw[<-,dashed] (2,2) -- (2,3) node [midway,fill=white] {NAKx};
\draw[<-] (3,1) -- (3,2) node [midway,fill=white] {$E_{\text{Tx}}^{\text{h}}$};
\draw[<-,dashed] (3,2) -- (3,3) node [midway,fill=white] {ACK};
\draw[<-] (4,1) -- (4,2) node [midway,fill=white] {$E_{\text{Tx}}^{\text{h}}$};
\draw[<-,dashed] (4,2) -- (4,3) node [midway,fill=white] {NAKx};
\draw[<-] (5,1) -- (5,2) node [midway,fill=white] {$E_{\text{Tx}}^{\text{h}}$};
\draw[<-,dashed] (5,2) -- (5,3) node [midway,fill=white] {ACK};
\draw[densely dotted] (1,0) -- (1,-.5);
\draw[<->] (0,-.20) -- (1,-.20) node [midway,fill=white] {$T_s$};
\draw[<->] (0,-.40) -- (3,-.40) node [midway,fill=white] {$T_f$};
\draw[densely dotted] (0,0) -- (0,-.5) node[below] {$t-5$};
\draw[densely dotted] (1,0) -- (1,-.5) node[below] {$t-4$};
\draw[densely dotted] (2,0) -- (2,-.5) node[below] {$t-3$};
\draw[densely dotted] (3,0) -- (3,-.5) node[below] {$t-2$};
\draw[densely dotted] (4,0) -- (4,-.5) node[below] {$t-1$};
\draw[densely dotted] (5,0) -- (5,-.5) node[below] {$t$};
\draw[densely dotted] (6,0) -- (6,-.5) node[below] {$t+1$};
\draw[<->] (0,.65) -- (1,.65) node [midway,above] {$E_{\text{Tx}}$};
\draw[<->] (1,.65) -- (1.75,.65) node [midway,above] {$E_{\text{Tx}}$};
\draw[<->] (2,.65) -- (2.25,.65) node [midway,above] {$E_{\text{Tx}}$};
\draw[<->] (3,.65) -- (4,.65) node [midway,above] {$E_{\text{Tx}}$};
\draw[-] (0,0) -- (1,0) node [midway,above] {$k=1$};
\draw[-] (1,0) -- (2,0) node [midway,above] {$k=2$};
\draw[-] (2,0) -- (3,0) node [midway,above] {$k=3$};
\draw[-] (3,0) -- (4,0) node [midway,above] {$k=1$};
\draw[-] (4,0) -- (5,0) node [midway,above] {$k=2$};
\node[text width=3cm] at (8.5,-.250) {time index};
\end{tikzpicture}

\caption{Time-slotted packet transmission time line at the transmitter node. '$\protect\dasharrow$'
and '$\protect\longrightarrow$' denote the \ac{ARQ} message and \ac{EH} arrival events, respectively. Shaded areas in the slot denote the amount of energy used for transmission.}
\label{fig:TransmissionTimeline}
\end{figure}
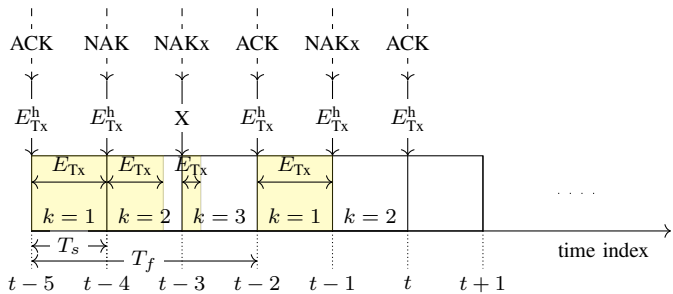

\section{System Model}
\subsection{\textcolor{black}{Transmission Model}}
We consider a point-to-point communication between two \ac{EH} wireless sensor nodes. Sensor nodes have limited capacity rechargeable batteries, which are charged by renewable energy sources. In \textcolor{black}{the considered} model, when a transmitted packet is erroneously decoded, the receiver requests its retransmission. A maximum of $K\in\mathbb{Z}$ \textcolor{black}{retransmission} requests are permitted. A packet consists of $c$ information bits, taken from the data buffer, encoded with an ($m,c$) \ac{FEC} code (e.g., convolutional code), and then modulated through an $M$-ary quadrature amplitude modulation, where $M$ denotes the cardinality of the constellation. This forms the packet of length \textcolor{black}{$\ceil{m/\log_{2}M}$} symbols, \textcolor{black}{ where $\ceil{\cdot}$ is the ceiling operator}.

The \ac{CSI} and \ac{SSI} are known at the transmitter through the \ac{ARQ} feedback messages. Each sensor node is aware of its own \ac{BSI}, but not \textcolor{black}{of the} \ac{BSI} of the other node. However, the transmitter estimates the one slot delayed \ac{BSI} of the receiver via \ac{SSI}.

A discrete time-slotted model is considered as depicted in Fig. \ref{fig:TransmissionTimeline}. Each time slot is of $T_{s}$ seconds duration and indexed as $t\text{\ensuremath{\in}\{1, 2,\ensuremath{\ldots}\}}$. The packet transmission and corresponding \ac{ARQ} message reception are completed within a slot, i.e., the round-trip time is $T_s$. Several slots constitute a frame of duration $T_{f}$. A frame has variable duration depending upon the number of slots being used in packet transmission, including retransmissions. Thus, the minimum and maximum \textcolor{black}{values} of frame duration \textcolor{black}{are} $T_s$ and $KT_s$, respectively. After a maximum of $K$ unsuccessful attempts, the transmitter drops the packet \textcolor{black}{and} chooses a new one to transmit.

\textcolor{black}{Without loss of generality, the system model considered here can be extended to generic short-range communication systems involving different modulation formats, sophisticated channel coding methods, and transmission strategies relying on multiple antennas and sub-carriers techniques.}

\subsection{Energy Consumption Model}
The transmitter and receiver sensor nodes \textcolor{black}{spend} energy to transmit and retrieve the information bits, respectively. For short-range commmunication, the energy consumption in a wireless link can be broken down into two dominant factors \cite{Cui-Goldsmith-Bahai-jsac-04}: the energy consumed at the power amplifiers $P_{\mathrm{PA}}$ at the transmitter, and the circuit blocks at both transmitter and receiver. \textcolor{black}{The} circuit blocks of the transmitter consist of a digital-to-analog convertor, mixers, active filters, and frequency synthesizers, while mainly of a low noise amplifier, intermediate frequency amplifier, active filters, analog-to-digital convertor, and frequency synthesizer at the receiver. Further, for coded systems, the energy expended in the decoding operation needs to be included \cite{Grover-Woyach-Sahai-jsac-11,Cui-Goldsmith-Bahai-jsac-04} at the receiver. Thus, for a coded system, the total energy expenditure at time slot $t$, at \textcolor{black}{both} transmitter and receiver nodes \textcolor{black}{is}, respectively, given as
\begin{alignat}{1}
P_{\mathrm{Tx}} & =\underbrace{(1+\alpha)P_{\mathrm{out}}}_{P_{\mathrm{PA}}}+P_{\mathrm{C,Tx}},\label{eq:TxEnergyConsumption}\\
P_{\mathrm{Rx}} & =P_{\mathrm{dec}}+P_{\mathrm{C,Rx}}+P_{\mathrm{fb}},\label{eq:RxEnergyConsumption}
\end{alignat}
where $P_{\mathrm{out}}$ is the transmit power, $\alpha=(\xi/\eta)-1$, with $\eta$ as the drain efficiency and $\xi$ as the peak-to-average power ratio. $P_{\mathrm{C,Tx}}$ and $P_{\mathrm{C,Rx}}$ are the total power spent in the circuit blocks of   the transmitter and receiver, respectively. The power consumed in transmitting the \ac{ARQ} messages is denoted by $P_{\mathrm{fb}}$.  $P_{\mathrm{dec}}$ denotes the power used in the decoding operation \textcolor{black}{and} is ignored for the uncoded system. Typical values of $P_{\mathrm{dec}}$ are around 70-80\% of the power dissipated in the circuit blocks \cite{Grover-Woyach-Sahai-jsac-11}. The index $t$ is dropped in (\ref{eq:TxEnergyConsumption}) and (\ref{eq:RxEnergyConsumption}) to simplify the presentation.

\subsection{Energy Harvesting Model}
The transmitter and receiver nodes are connected to two separate but similar renewable \ac{EH} sources. In particular, two \ac{i.i.d.}, Bernoulli \textcolor{black}{random processes} are considered to model the energy \textcolor{black}{arrivals,} similar to \cite{Paradiso-Feldmeier-icuc-09}. The Bernoulli model is tractable and captures the intermittent and irregular behavior of the energy arrival. It is worth mentioning that this work is, in essence, independent of the energy arrival process; this will be shown later in the simulation results, where \textcolor{black}{the} compound Poisson arrival model \cite{Xu-Zhang-jsac-14} is used \textcolor{black}{as well}. \textcolor{black}{At the start of every slot, $E^{\mathrm{h}}_{\mathrm{Tx}}$ Joule (J) with probability (w.p.) $\rho_{\text{Tx}}$ and zero J w.p.  $1-\rho_{\text{Tx}}$ is harvested at the transmitter. The receiver node follows a similar energy arrival process with probability $\rho_{\text{Rx}}$ and amount $E^{\mathrm{h}}_{\mathrm{Rx}}$.}

\textcolor{black}{When the two nodes are in close vicinity and have the same type of harvesting source, then the two \ac{EH} processes are spatially correlated. In this case, the harvested energy pairs at time slot $t$ are given as \cite{Zhou-Chen-Chen-Niu-jsac-15}
\begin{IEEEeqnarray*}{lCr}
(E^{t}_{\mathrm{Tx}}, E^{t}_{\mathrm{Rx}})=
\begin{cases}
(0,0) & \small{\text{w.p.\,\,}p_{00}},\\
(0,E^{\mathrm{h}}_{\mathrm{Rx}}) & \small{\text{w.p.\,\,}p_{01}},\\
(E^{\mathrm{h}}_{\mathrm{Tx}},0) & \small{\text{w.p.\,\,}p_{10}},\\
(E^{\mathrm{h}}_{\mathrm{Tx}},E^{\mathrm{h}}_{\mathrm{Rx}}) & \small{\text{w.p.\,\,}p_{11}},\\
\end{cases}\IEEEyesnumber\label{eq:BatteryEvolution}
\end{IEEEeqnarray*}
with the condition that $p_{00}+p_{01}+p_{10}+p_{11}=1$. For example, when $p_{00}=p_{01}=p_{00}=p_{10}=p_{11}=0.25$ then both nodes harvest energies independent from each other. When $p_{01}=p_{10}=0$ and $p_{00}=p_{11}=0.5$, the harvested energies are highly correlated.}

Let $B_{\mathrm{Tx}}^{t}$ and $B_{\mathrm{Rx}}^{t}$ denote the energy levels of the battery at the start of the $t$th time slot, and $E_{\mathrm{Tx}}^{t}=P_{\mathrm{Tx}}T_{s}$ and $E_{\mathrm{Rx}}^{t}=P_{\mathrm{Rx}}T_{s}$ denote the energy consumed in transmitting a packet, \textcolor{black}{as well as} sampling and decoding operations at the transmitter and receiver, respectively. The battery level at the transmitter follows the Markovian evolution:

\begin{IEEEeqnarray*}{lCr}
B_{\mathrm{Tx}}^{t+1}=
\begin{cases}
\min\{B_{\mathrm{Tx}}^{t}+E^{\mathrm{h}}_{\mathrm{Tx}}-E_{\mathrm{Tx}}^{t},B^{\mathrm{max}}_{\mathrm{Tx}}\}, & \small{\text{w.p.\,\,}\textcolor{black}{\rho_{\text{Tx}}}}\\
B_{\mathrm{Tx}}^{t}-E_{\mathrm{Tx}}^{t}, & \small{\text{w.p.\,\,} 1-\textcolor{black}{\rho_{\text{Tx}}},}
\end{cases}\IEEEyesnumber\label{eq:BatteryEvolution}
\end{IEEEeqnarray*}
where $B^{\mathrm{max}}_{\mathrm{Tx}}$ denotes the transmitter node's battery capacity. Replacing the subscript $\mathrm{Tx}$ in (\ref{eq:BatteryEvolution}) with $\mathrm{Rx}$ gives the receiver side battery evolution. For presentation simplicity, the energies are normalized by a minimum possible energy, i.e., $E^{\mathrm{min}}_{\mathrm{Tx}}$ and $E^{\mathrm{min}}_{\mathrm{Rx}}$ that are \textcolor{black}{spent} in transmitting and receiving a packet of smallest size, respectively. Consequently, the transmitter energy level is an integer multiple of $E^{\mathrm{min}}_{\mathrm{Tx}}$ and the change in the battery state whenever harvesting takes place is $L_{\mathrm{Tx}}\triangleq E^{\mathrm{h}}_{\mathrm{Tx}}/E^{\mathrm{min}}_{\mathrm{Tx}}$. Similarly, the battery energy level at the receiver side is an integer multiple of $E^{\mathrm{min}}_{\mathrm{Rx}}$ and the \ac{EH} amount is $L_{\mathrm{Rx}}\triangleq E^{\mathrm{h}}_{\mathrm{Rx}}/E^{\mathrm{min}}_{\mathrm{Rx}}$.

\subsection{Channel Model}
The wireless channel from the transmitter to the receiver is assumed to be Rayleigh faded and modeled as a \ac{FSMC} \cite{Wang-Moayeri-tvt-95,Zhang-Kassam-tcom-99}. This model captures the main features of fading channels, and approximates the fading as a discrete-time Markov process. Essentially, all possible fading gains are modelled as a set of finite and discrete channel states. The \ac{FSMC} channel is described as follows: discrete states of the channel $\mathcal{G}=\{g_{1,}g_{2},\ldots,g_{\textcolor{black}{|\mathcal{G}|}}\}$, state transition probabilities $\mathbf{\boldsymbol{\Omega}}=\{p(g_{j}|g_{i}):g_{1}<g_{i,}\, g_{j}<g_{\textcolor{black}{|\mathcal{G}|}}\}$, and steady state probabilities represented by $\omega_{o}(g_{i})$, $i=1,2,\ldots,\textcolor{black}{|\mathcal{G}|}$.

Based on the \ac{FSMC} channel model, the entire range of channel gains are partitioned into $\textcolor{black}{|\mathcal{G}|}+1$ non-overlapping intervals with boundary values denoted as $\{\gamma_i\}^{\textcolor{black}{|\mathcal{G}|}}_{i=0}$, with increasing order of their values from $\gamma_{0}=0$ to $\gamma_{\textcolor{black}{|\mathcal{G}|}}=\infty$. The fading gain interval $[\gamma_{i-1},\gamma_{i})$ represents the $g_{i}$ channel state\textcolor{black}{, which is considered fixed during the time slot $t$ and changes to $g_{j}$ in time slot $t+1$ with probability $p(g_{j}|g_{i})$.}
The channel state is considered fixed during time slot $t$ and changes to another in time slot $t+1$ with probability $p(g_{j}|g_{i})$ .

\begin{table}[h]
\renewcommand{\arraystretch}{0.40}
\centering
\textcolor{black}{\caption{List of symbols}
\label{tab:symbolslist}
\centering
\begin{tabular}{c|p{5.2cm} ||c|p{5.2cm}}
\hline
\bfseries Symbol & \bfseries Description & \bfseries Symbol & \bfseries Description\\
\hline
\hline
$K$ and $k$ & Maximum number of retransmissions and its index 																													& $\mathcal{S}$, $|\mathcal{S}|$ and $S_t $ & System state space, its cardinality, and state at time slot $t$\\
\hline
$c$ and $m$ & Number of information and coded bits 																																& $\mathcal{B}_{\text{Tx}}$, $\mathcal{B}_{\text{Rx}}$, $|\mathcal{B}_{\text{Tx}}|$, and $|\mathcal{B}_{\text{Rx}}|$ & Transmitter and receiver battery states space and their cardinality\\
\hline
$M$ and $R_c$ & Modulation order and code rate  															    																& $\mathcal{U}_{S_t}$, $|\mathcal{U}_{S_t}|$, $a_t$, $B_{S_t}$ & Action states space, its cardinality, action, and maximum value of an action at time slot $t$  \\
\hline
$T_s$ and $T_f$ & Slot duration and frame duration 																																&  $\mathcal{Z}$, $|\mathcal{Z}|$ and $Z_t$ & Observation state space, its cardinality, and observation state at  time slot $t$ \\
\hline
$E^t_{\text{Tx}}$ and $E^t_{\text{Rx}}$ & Transmitter and receiver energy expenditure at time slot $t$ 																			& $\text{ACK, NAK}$, $\text{NAKx}$ & ARQ feedback messages \\
\hline
$P_{\text{dec}}$ and $P_{\text{fb}}$ & Decoding and ARQ message transmit power expenditure at receiver 																			& $P_e(g,a_t)$ & Packet error probability with energy $aE^{\text{min}}_{\text{Tx}}$ via channel state $g$\\
\hline
$P_{\text{out}}$ and $P_{\text{PA}}$& Transmit power and power amplifier output power 																							& $P_2(d,g,a_t)$ & Modulation dependent bit error probability\\
\hline
$P_{\text{C,Tx}}$ and $P_{\text{C,Rx}}$ & Transmitter and receiver circuit block power  																						& $P_{\text{err}}$ & Approximate packet error probability after NAK or decoding failure \\
\hline
$\eta$ and $\xi$ & Amplifier drain efficiency and PAPR 																															& $A_d$ and $d_{\text{free}}$ & Weight spectral coefficient and free distance of convolution code\\
\hline
$B^{\text{max}}_{\text{Tx}}$ and $B^{\text{max}}_{\text{Rx}}$ & Maximum battery size of transmitter and receiver nodes 															& $r(s,a_t)$ & Cost function of state $S_t = s$ after taking action $a$ \\
\hline
$E^{h}_{\text{Tx}}$ and $E^{h}_{\text{Rx}}$ & Transmitter and receiver EH amounts   																							& $\rho_{\text{Tx}}$ and $\rho_{\text{Rx}}$ & Transmitter and receiver nodes probabilities of EH   \\
\hline
$E^{\text{min}}_{\text{Tx}}$ and $E^{\text{min}}_{\text{Rx}}$ & Minimum energy to transmit and receive a packet of minimum size 												& $\pi$ and $J_{\pi}(S_0)$ & Transmitter policy and total expected cost with given start state $S_0$   \\
\hline
$L_{\text{Tx}}$ and $L_{\text{Rx}}$ & $E^h_{\text{Tx}}/E^{\text{min}}_{\text{Tx}}$ and $E^h_{\text{Rx}}/E^{\text{min}}_{\text{Rx}}$ 											& $\varpi(G_{t})$ & Belief of channel state $G_t$ at time slot $t$  \\
\hline
$\mathcal{G}$, $|\mathcal{G}|$, $\gamma_i$, and $G_t$ & Total number of discrete channel states, its cardinality, $i$th interval fading gain, and channel state at time slot $t$ & $P_{\text{drop}}$ and $\bar{P}_{\text{drop}}$ & PDP and average PDP \\
\hline
$g_i$ and $\omega_o(g_i)$  & Channel state of interval $[\gamma_{i-1}, \gamma_i)$ and its steady state probability 																& $\psi(i,j)$ & Stationary probability  distribution with transmitter and receiver energies $(iE^{\text{min}}_{\text{Tx}}, jE^{\text{min}}_{\text{Rx}})$  \\
\hline
$\beta$ and $\text{x}$ & Number of division of a packet and number of additional ARQ messages 																					& $\Xi^{q,r,w,y}_{i,j,z,k}$ & Transition probability of going from state $(i,j,z,k)$ to state $(q,r,w,y)$ \\
\hline
\end{tabular}}
\end{table}

\section{Adaptive Retransmission Scheme}
In low harvesting rate, the receiver might not perform the sampling and decoding operations together or perform only the sampling operation in one time slot. In such time slots, the receiver performs \ac{SS} and stores the samples\textcolor{black}{, which can be later combined with the remaining parts of the packet for building a full packet.} Besides, the receiver sends back the \ac{SSI} via \ac{ARQ} messages. The transmitter then adapts the packet length for the next transmission. In the adaptive retransmission scheme, \ac{ARQ} messages carry $1 + \log_{2}\beta$ bits, where $\beta\in\{1,2,4,8,\ldots,\textcolor{black}{\ceil{m/\log_{2}M}}\}$, and have a total of $\mathrm{x}+2$ messages where
\begin{flalign}
\mathrm{x}\in & \begin{cases}
{\{0,1,\ldots,\beta\}}, & \beta>1\\
\emptyset, & \mathrm{otherwise.}
\end{cases}\label{eq:ACK/NAKxScheme}
\end{flalign}
For $\beta=1$, the scheme becomes the conventional one. The additional $\mathrm{x}$ messages are essentially the \ac{SSI} which are carried back to the transmitter as \ac{ARQ} messages; henceforth, we refer to the adaptive retransmission scheme by \ac{ACK}/NAKx. Details of each message are as follows:

\subsubsection*{\noindent $\mathrm{ACK}$}

The packet decoding at the receiver is successful. In reply, the transmitter chooses a new packet for transmission in the subsequent time slot.

\subsubsection*{\noindent $\mathrm{NAK}$}

The packet decoding is erroneous. In reply, the transmitter chooses the same packet for transmission in the subsequent time slot.

\subsubsection*{\noindent $\mathrm{NAKx}$}

$\textcolor{black}{\ceil{\mathrm{x}m/(\beta\log_{2}M)}}$ symbols of the transmitted packet are sampled and the rest is discarded due to the lack of energy. In reply, in the subsequent time slot, the transmitter sends a packet with the remaining $\textcolor{black}{\ceil{m(\beta-\mathrm{x})/(\beta\log_{2}M)}}$ symbols.

Note that the message \ac{NAK} is different from NAK0, which corresponds to the case when the receiver does not have enough energy to sample the smallest fraction of the transmitted packet. However, in both cases, the transmitter retransmits the full packet. As for the conventional scheme, the \ac{ACK}/NAKx messages \textcolor{black}{help in estimating the} \ac{CSI} to the transmitter node. Furthermore, it is now evident that the SS is a function of the available energy \textcolor{black}{at the receiver}. \textcolor{black}{Based on the chosen value of $\beta$, the receiver selects $\ceil{\mathrm{x}m/(\beta\log_2M)}$ symbols for sampling in any given time slot, where $\mathrm{x}$ is related to the available energy at the receiver. The transmitter sends the $\ceil{m(\beta-\mathrm{x})/(\beta\log_{2}M)}$ symbols after receiving NAKx message from the receiver. The length of the transmitted packet depends on the energy available at the transmitter. For example, if $\beta = 4$,  the variable $\mathrm{x}$ takes values from $\{0,1,2,3,4\}$. For $\mathrm{x} = 3$ and if $B_{\mathrm{Rx}}\geq \mathrm{x}E_{\mathrm{C,Rx}}/\beta = 3E_{\mathrm{C,Rx}}/4$, the receiver samples $\ceil{\mathrm{x}m/(\beta\log_2M)} = \ceil{0.75m/\log_2M}$ symbols and selects NAK\footnotesize{3} \normalsize to feedback to the transmitter. At the transmitter, if $B_{\mathrm{Tx}}\geq E^{\text{min}}_{\text{Tx}}$ then a part of the packet of size $\ceil{m(\beta-\mathrm{x})/(\beta\log_{2}M}) = \ceil{0.25m/\log_{2}M}$ symbols is sent out; otherwise, there is no transmission}. In another example, if $B_{\mathrm{Rx}}\geq (E_{\mathrm{C,Rx}}+E_{\mathrm{dec}})$, the receiver samples and decodes the packet and selects the \ac{NAK} or \ac{ACK} message depending on the decoding outcome. \textcolor{black}{In the conventional retransmission scheme, if the receiver lacks energy to sample the full packet, it samples the packet till the energy lasts and does not store the samples, and requests the packet retransmission, which requires full amount of energy. In conclusion, the conventional retransmission scheme wastes energy when compared with the proposed adaptive one.}

\section{Power Assignment Strategy}
Considering the adaptive retransmission scheme, we formulate in this section the power assignment as a sequential decision problem and then discuss two methods for \textcolor{black}{managing efficiently} the harvested energy at transmitter. At each time slot, the transmitter chooses the energy levels that minimize the average \ac{PEP}. The decision is based on the retransmission index, \ac{BSI}, sequences of past \textcolor{black}{observations} and power assignments at the transmitter. \textcolor{black}{After each transmission, the transmitter receives a feedback, referred to as observation $\{\mathrm{ACK,NAK,NAK\mathrm{x}}\}$, from the receiver. Furthermore, based on the observation, the transmitter can also adapt the modulation and coding scheme. However, for sake of tractability, we only consider the transmit packet size and power adaptations.} The problem is considered in infinite horizon.

\subsection{Problem Formulation}
We formulate the problem by defining the following components: a set of time slots $\mathcal{T}=\{1,2,\ldots\}$ over which decisions \textcolor{black}{are} made, and a set of system states $\mathcal{S}$, a set of transmitter \ac{BSI} $\mathcal{B}_{\mathrm{Tx}}$, a set of \ac{FSMC} channel states $\mathcal{G}$, a set of retransmission indices $\mathcal{K}=\{1,2,\ldots,K\}$, a set of actions $\mathcal{U}$, a set of transition probabilities $\mathcal{P}$, a set of observations $\mathcal{Z},$ and a cost corresponding to every decision. Let $\mathcal{S}=\mathcal{B}_{\mathrm{Tx}}\times\mathcal{G}\times\mathcal{K}=\{(b_{1},g_{1},k_{1}),(b_{2},g_{1},k_{1}),\ldots,(b_{|\mathcal{B}_{\text{Tx}}|},g_{|\mathcal{G}|},k_{|\mathcal{K}|})\}$ denote the complete discrete state space of the system with a total of \textcolor{black}{ $|\mathcal{B}_{\text{Tx}}|\times |\mathcal{G}|\times |\mathcal{K}|$} states, where $b$, $g$, and $k$ represent the transmitter battery state, channel state, and retransmission index state, respectively. \textcolor{black}{The state of the system, channel, and observation at time slot $t$ are represented as $S_t\in \mathcal{S}$, $G_t\in \mathcal{G}$, and $Z_t\in \mathcal{Z}$, respectively. The} retransmission index $k$ tracks the system state within each frame and is reset to one when its maximum value $K$ is reached or when \ac{ACK} is received, whichever comes first. \textcolor{black}{Due to \ac{EH}, the cardinality of the actions set varies in every time slot. Hence, the set of actions at time slot $t$ is denoted by} $\mathcal{U}_{s_{t}}\triangleq\{0,1,2,\ldots,B_{s_{t}}\}$, where $B_{s_{t}}\in\mathcal{\mathcal{B}_{\mathrm{Tx}}}$ represents the current state battery level. \textcolor{black}{This} is a set consisting of feasible choices of energy levels corresponding to the transmission of a full packet. An action $a_t\in\mathcal{U}_{s_{t}}$ \textcolor{black}{represents the energy level $a_tE^{\mathrm{min}}_{\mathrm{Tx}}=P_{\mathrm{out}}T_{s}$ in time slot $t$}. For each action taken, the system receives an observation \textcolor{black}{belonging to the set $\mathcal{Z}$. Note that a set of receiver \ac{BSI} can be included in the system state space. Since the exact receiver BSI state is unkown to the transmitter, its value can be estimated using the ARQ messages likewise the channel state. However, including more unknown states to the system state space increases the complexity in solving the problem.} 

\textcolor{black}{We first define the \ac{PEP} used for the proposed adaptive retransmission scheme. Note that the packet can be transmitted in parts, as presented in Section-III, and the receiver can decode the packet only when all the samples are available. Since the different parts of the packet have passed through different channel states, the part which has passed through the worse channel state leads to the decoding failure. To simplify, we assume that the full packet is transmitted through the worse channel state, and the receiver decodes it. Thus, the approximate \ac{PEP} is the one of the worse part. Consequently, we use the following approximate \ac{PEP} expression in the rest of the paper.}
\begin{definition}
For $\beta > 1$, the \ac{PEP} after the transmitter receives $Z_{t}=\mathrm{NAK}$ in current time slot $t$ and retransmission index $K_t=k$ is approximated as
\textcolor{black}{\begin{IEEEeqnarray}{rCl} \label{eq:ErrorProbDef}
P_\mathrm{err}(G_{t},a_{t})\approx \max_l\{P_{e}(G_{t-l},a_{t-k})\}, \qquad 1\leq l \leq k,
\end{IEEEeqnarray}
where $P_{e}(G_t,a_t)$ is the probability that a full packet transmitted in time slot $t$, with energy $a_t E^{\mathrm{min}}_{\mathrm{Tx}}$ via channel state $G_t=g$, received in error. $a_{t-k}$ corresponds to the energy level used by the transmitter when sending a new packet at the retransmission $k=1$. Through numerical simulations, we have verified that the approximated \ac{PEP} approaches the simulated \ac{PEP}; hence, the approximation \eqref{eq:ErrorProbDef} is reliable. Furthermore, for $\beta = 1$ the above expression becomes same as $P_e(G_t,a_t)$. }
\end{definition}

Furthermore, \ac{PEP} is a function of the modulation type and \ac{FEC} coding used. With the convolution code, for example, \ac{PEP} is calculated as \cite{Pursley-Taipalel-tcom-87}:
\begin{equation}
P_{e}\big(g,a\big) \leq1-\Big(1-{\displaystyle \sum_{d=d_{\mathrm{free}}}^{m}A_{d}P_{2}(d,g,a)}\Big)^{m},
\label{eq:PacketErrorProb}
\end{equation}
where $d_{\mathrm{free}}$ is the free distance and $A_{d}$ is the weight spectra coefficients of the convolutional code. $P_{2}(d,g,a)$ is the modulation dependent bit error probability. For example, the \acl{BEP} of binary phase-\textcolor{black}{shift}-keying can be approximated by $P_{2}(d,g,a)\approx 0.5\mathrm{erfc}(\sqrt{(d\tilde{\gamma}(g)P_{\mathrm{out}})/\sigma_{n}^{2}})$, where $\sigma_{n}^{2}$ is the noise power,
$\mathrm{erfc(\cdot)}$ is the complementary error function, and $\tilde{\gamma}(g)$ is the average power gain in channel state $g$, which can be found as $\tilde{\gamma}(g)=(\int_{\gamma_{i-1}}^{\gamma_{i}}\gamma p(\gamma)d\gamma)/\int_{\gamma_{i-1}}^{\gamma_{i}}p(\gamma)d\gamma$, \textcolor{black}{where $p(\gamma)$ is the probability density function of $\gamma$, which is distributed exponentially.} Here, we assume that the error detection code is able to find all remaining errors. 

Now, we consider a system state at time $t$ as $S_{t}=(B^{t}_{\mathrm{Tx}}=b,G_{t}=g,K_{t}=k)$. Let the \ac{ARQ} message at time $t$ be denoted by $Z_{t}$, where $Z_{t}=\mathrm{NAK}$ for decoding failure, $\mathrm{NAK\text{x}}$ for incomplete transmission or $Z_{t}=\mathrm{ACK}$ for a decoding success. After an action $a_{t}$ is taken at time slot $t$, the current system state goes to a new state with the transition probability $p(S_{t+1}=s^{\prime}|S_{t}=s,a_{t})$ and \textcolor{black}{is associated with} a cost. Let $s=(b,g,k)$ be the current system state, then $r(S_{t}=s,a_{t})$ is the cost defined as:
\begin{gather}
r(s,a_t)=\begin{cases}
P_{\mathrm{err}}(g,a_t) & a_t \leq b,Z_{t}=\mathrm{NAK},\\
0 & \mathrm{otherwise}.
\end{cases}\label{eq:RewardFunc-1}
\end{gather} 
The cost function is independent of the receiver available energy as it is unknown at the transmitter. 

At time slot $t$, the probability of transition from state $s=(b,g,k)$ to state $s^{\prime}=(b^{\prime},g^{\prime},k^{\prime})$ after taking an action $a_t$, similar to \cite{Aprem-Murthy-Mehta-jstsp-13}, is given as
\begin{equation}
p(s^{\prime}|s,a_t)=\delta(k^{\prime},k_{+})p(g^{\prime}|g)\zeta((b^{\prime},a_t,b,k,g),\label{eq:Tx_trans_prob_def}
\end{equation}
where $k_{+}\triangleq(k\enspace\mathrm{mod}\enspace K)\boldsymbol{1}_{Z_{t+1}\neq\mathrm{ACK}}+1$, with the indicator function $\boldsymbol{1}_{A}$ equal \textcolor{black}{to 1 if the event $A$ is true, and to zero otherwise.} $\delta(\cdot,\cdot)$ is the Kronecker delta function. $\delta(k^{\prime},k_{+})$ ensures that the transmission index increases by one at each state transition and is reset to one when the maximum retransmission times is reached or when $Z_{t+1}=$ \ac{ACK} is received, whichever occurs first. $\zeta(b^{\prime},a_t,b,k,g)$ is the probability that the transmitter with current channel \textcolor{black}{state} and retransmission state $(g,k)$ moves from battery state $b$ to another state $b^{\prime}$ after taking an action $a_t$. For $k\geq1$,
\begin{IEEEeqnarray*}{lCl}
\zeta(b^{\prime},a_t,b,k,g)=  \eta(b^{\prime},a_t,b) 
\times\begin{cases}
\textcolor{black}{\rho_{\text{Rx}}} P_{\mathrm{err}}(g,a_t) & \small{Z_{t+1}=\mathrm{\mbox{NAK},}Z_{t}=\mathrm{\mbox{NAK,\,NAKx}}}\\
\textcolor{black}{\rho_{\text{Rx}}} (1-P_{\mathrm{err}}(g,a_t)) & \small{Z_{t+1}=\mathrm{\mbox{ACK},}Z_{t}=\mathrm{\mbox{NAK,\,NAKx}}}\\
1-\textcolor{black}{\rho_{\text{Rx}}} & \small{Z_{t+1}=\mbox{NAKx},Z_{t}=\mathrm{\mbox{NAK,\,NAKx}}}\\
0 & \small{\mathrm{otherwise}},
\end{cases}
\IEEEyesnumber \label{eq:Tx_trans_prob_detail}
\end{IEEEeqnarray*}
where $\eta(b^{\prime},a_t,b)\triangleq\textcolor{black}{\rho_{\text{Tx}}}\text{\ensuremath{\delta}}(b^{\prime},b+L_{\mathrm{Tx}}-a_t)+(1-\textcolor{black}{\rho_{\text{Tx}}})\text{\ensuremath{\delta}}(b^{\prime},b-a_t)$.

At time slot $t$, the transmitter uses the history of both \textcolor{black}{observation} sequence, i.e., $\mathbf{z}{}_{t}\triangleq[Z_{1},\ldots,Z_{t}]$ with $Z_{1}=$ \ac{ACK}, and previously selected transmit power $\mathbf{a}_{t-1}\triangleq[a_{1},\ldots,a_{t-1}]$ to choose the transmit power $a_{t}$ from the set $\mathcal{U}_{S_t}$ of the admissible
power level. The transmit power is selected to minimize the total expected cost for the current and remaining packets:
\begin{IEEEeqnarray*}{lCr}
a^{\star}_t & \triangleq{\displaystyle {\displaystyle \arg\min_{a_t{\in\mathcal{U}_{\textcolor{black}{S_{t}}}}}\mathbb{E}}\bigg\{ r(S_{t},a_t)+\sum_{k=t+1}^{\infty}r(S_{k},a^{\star}_k)\bigg|\mathbf{z}{}_{t},}\mathbf{a}_{t-1}\bigg\}\\
 & \mathrm{for}\hspace{1em}t=1,2,\ldots,
\IEEEyesnumber \label{eq:TotalExpProbError}
\end{IEEEeqnarray*}
\textcolor{black}{where $a^{\star}_t$ and $a^{\star}_k$ denote the optimal transmit power assignments for  the current time slots $t$ and for future time slot $k=t+1$, respectively. $\mathbb{E}\{\cdot\}$ is the expectation operator.}

\textcolor{black}{Let the policy $\pi:\mathcal{S}\rightarrow\mathcal{U}$ specifies the rule for the selection of an action by the transmitter in a given time slot. Hence, a policy is basically a mapping between what happened in the past and what has to be done at the current state. To find a policy $\pi$ that minimizes the total expected cost, we cast \eqref{eq:TotalExpProbError} as an infinite-horizon \ac{MDP} as}
\begin{equation}
J_{\pi}(S_{0})=\lim_{T\rightarrow\infty}\frac{1}{T}\mathrm{\mathbb{E}}\bigg\{\sum_{t=1}^{T}r(S_{t},a_t)\bigg|S_{0}, \mathbf{z}_{t},\mathbf{a}_{t-1}\bigg\},\label{eq:LongTermAvgReward}
\end{equation}
where $S_{0}$ is a known start state. The optimal policy $\pi^{\star}$ minimizes the expected long-term average cost given by (\ref{eq:LongTermAvgReward}). \textcolor{black}{The optimal policies obtained in infinite-horizon \ac{MDP} problems are often stationary, and hence, simpler to implement compared to what is obtained in a finite-horizon \ac{MDP} problem that varies in each time slot. Furthermore, since the total system state space is countable and discrete, and $\mathcal{U}_{S_t}$ is finite for each $S_t \in \mathcal{S}$, there exists an optimal stationary deterministic policy $\pi^{\star}$ that minimized the total expected cost.}

\subsection{Solution Methods}
\textcolor{black}{This section discusses the solution methods for solving the \ac{MDP} considered in this work.} The following Bellman equation \cite{Bertsekas-01} is used to
solve \textcolor{black}{\eqref{eq:LongTermAvgReward},}
\begin{equation}
\lambda^{\star}+h^{\star}(s)=\min_{a_t\in\mathcal{U}_{s},a_t\leq B_{s}}\Big[r(s,a_t)+\sum_{s^{\prime}\in\mathcal{S}}p(s^{\prime}|s,a_t)h^{\star}(s^{\prime})\Big],\label{eq:BellmanEq}
\end{equation}
where $\lambda^{\star}$ is the optimal cost and $h^{\star}(s)$ is an optimal differential cost or relative value function for each state $s \in \mathcal{S}$. \textcolor{black}{The Bellman equation is a well estabilished and commonly used method for solving a sequential decision making problem. Interested readers may refer to \cite{Putterman-94,Bertsekas-01} for further insights.} Let $\pi^{\star}(s)$ denote the solution of the \ac{MDP} solved via the value iteration algorithm.

In the formulation described above, the exact \ac{CSI} is unknown at the transmitter while making the decisions. Since one of the system state variable is partially known, the problem at hand is commonly referred to as \ac{POMDP} \cite{Putterman-94}. Consequently, based on the observation history, a belief channel state space of the system is formed. It represents a sufficient statistic for the history of the previous actions and observations, and adequate actions can be chosen depending upon the belief state. The belief channel state $\varpi(G_{t})=p(G_{t}|\textbf{z}_{t},\textbf{a}_{t-1})$ is defined as a probability distribution over all possible states conditioned on the history of previous actions and observations. The belief state at time slot $t$ can be obtained by expanding the inferred \ac{CSI} distribution via \textcolor{black}{the} Bayes rule

\begin{IEEEeqnarray*}{lCr}
\varpi(G_{t})=\sum_{j=1}^{G} p(G_{t}|G_{t-1}=g_{j},\textbf{z}_{t},\textbf{a}_{t-1})p(G_{t-1}=g_{j}|\textbf{z}_{t},\textbf{a}_{t-1})\\
{\displaystyle =\sum_{j=1}^{G}p(G_{t}|G_{t-1}=g_{j})p(G_{t-1}=g_{j}|\textbf{z}_{t},\textbf{a}_{t-1})},\IEEEyesnumber \label{eq:CSIdistribution1}
\end{IEEEeqnarray*}
where we use the Markov \ac{CSI} variation assumption to write (\ref{eq:CSIdistribution1}). Further, with some simple mathematical manipulations, (\ref{eq:CSIdistribution1}) can be written as:
\begin{IEEEeqnarray*}{lCr}
p(G_{t-1}|\textbf{z}_{t},\textbf{a}_{t-1})= 
\frac{p(Z_{t-1}|a_{t-1},G_{t-1})p(G_{t-1}|\textbf{z}_{t-1},\textbf{a}_{t-2})}{\sum_{l=1}^{G}p(Z_{t-1}|a_{t-1},G_{t-1}=g_{l})p(G_{t-1}=g_{l}|\mathbf{z}_{t-1},\mathbf{a}_{t-2})},\qquad \IEEEyesnumber \label{eq:CSIdistribution2}
\end{IEEEeqnarray*}
where for $G_{t}=g$
\begin{IEEEeqnarray*}{lCr}
p(Z_{t}|a_t,g)=
\begin{cases}
\textcolor{black}{\rho_{\text{Rx}}} P_{\mathrm{err}}(g,a_t) & \small{Z_{t}=\mbox{NAK},a_t>0,}\\
\textcolor{black}{\rho_{\text{Rx}}} (1-P_{\mathrm{err}}(g,a_t)) & \small{Z_{t}=\mbox{ACK},a_t>0},\\
1-\textcolor{black}{\rho_{\text{Rx}}} & \small{Z_{t}=\mbox{NAKx},a_t>0},\\
p(Z_{t-1}|a_t,g) & \small{a_t=0}.
\end{cases}
\IEEEyesnumber \label{eq:ErrorRateProb}
\end{IEEEeqnarray*}
Note that when $a_t=0$, the transmitter is \textcolor{black}{in energy outage}; thus, no transmission takes place. \textcolor{black}{Consequently}, the acknowledged state also remains the same as the previously received one and so is the probability.

The \ac{POMDP} can be solved using dynamic programming to find the optimal policy. \textcolor{black}{However, solving the POMDP optimally is computationally infeasible for systems with a total number of states higher than 15 \cite{Littman-icml-95}. In our case, the total number of system states is large, and hence, the optimal solution is not presented.} \acp{POMDP} are PSPACE-complete, i.e., they have high computational complexity and require large memory that grows exponentially with the horizon \cite{Papadimitrious-Tsitsiklis-mor-87}. Furthermore, PSPACE-complete problems are even harder than NP-complete problems. However, many heuristics exist to find suboptimal policies, e.g., \ac{MLPH} \cite{Nourbakhsh-Powers-Birchfield-aimag-95}.

\subsection{Proposed Solutions}

\subsubsection{MLPH Power Assignment}
We solve \textcolor{black}{the} problem \textcolor{black}{in \eqref{eq:BellmanEq}} using the \ac{MLPH} method. In this approach, we first determine the state that the channel is most likely in, i.e.,
\begin{equation}
g_{\mathrm{ML}}=\arg\max_{G_{t}\in\mathcal{G}}\varpi(G_{t}).\label{eq:ML_estimate}
\end{equation}

With $\gamma_{\mathrm{ML}}$ \textcolor{black}{as} the belief channel state at \textcolor{black}{the} $t$th time slot, the corresponding ML state is denoted as $s_{\mathrm{ML}}=(b,g,k).$ Then, the transmit power policy is set as
\begin{equation}
a_t\triangleq\pi^{\star}(s_{\mathrm{ML}}).
\end{equation}

Furthermore, \ac{MLPH} finds the most probable state of the system from the belief state. When two or more states are equally likely, \ac{MLPH} chooses one arbitrarily.

\subsubsection{Greedy Power Assignment}
For low-power wireless sensors, the optimal solution should be avoided due to \textcolor{black}{the} computational complexity constraint. Thus, we turn to a suboptimal greedy power assignment scheme by modifying (\ref{eq:TotalExpProbError}) as:

\begin{equation}
\begin{array}{cl}
\bar{a}_t\triangleq & \arg{\displaystyle \min_{a_t\in\mathcal{U}_{S_{t}}}\mathbb{E}\{r(s,a_t)|\mathbf{z}_{t},\mathbf{a}_{t-1}\}}\\
 & \text{for}\quad \mathbf{ }t=1,2,\ldots.
\end{array}.\label{eq:GreedyPowerAlloc}
\end{equation}
\textcolor{black}{The main idea of greedy power assignment is to avoid computation of future dependent expected cost values. This incurs performance loss, however, at the expense of lower computational and storage requirement.} The greedy power assignment scheme can be rewritten as:

\begin{equation}
\bar{a}_t=\arg\min_{a_t\in\mathcal{U}_{S_{t}}}{\displaystyle \sum_{i=1}^{G}r(s,a_t)p(G_{t}=g_{i}|\mathbf{z}_{t},\mathbf{a}_{t-1})}.\label{eq:ExpandGreedyPowerAlloc}
\end{equation}

In order to estimate the greedy power assignment, (\ref{eq:ExpandGreedyPowerAlloc}) has to be implemented recursively. Using (\ref{eq:CSIdistribution1}) and (\ref{eq:CSIdistribution2}), we can write the following recursive implementation for \textcolor{black}{the} greedy power assignment.
\begin{enumerate}
\item \noindent Measure $Z_{t}$, compute $p(Z_{t}|a_{t-1},G_{t-1})$
as a function of $G_{t-1}$, and calculate $p(G_{t}|\mathbf{z}_{t},\mathbf{a}_{t-1})$
using (\ref{eq:CSIdistribution2}).
\item \noindent Calculate $p(G_{t}|\mathbf{z}_{t},\mathbf{a}_{t-1})$ using
the Markov prediction step (\ref{eq:CSIdistribution1}).
\item \noindent Calculate $a_t$ via (\ref{eq:GreedyPowerAlloc}).
\end{enumerate}
For the initial packets indices $t\in\{1,2\}$, we use the initial steady state distribution of states $\omega_{o}$ \textcolor{black}{instead} of
$p(G_{t-1}|\mathbf{z}_{t-1},\mathbf{a}_{t-2})$.

\subsubsection{Implementation Issues}
Here, we compare \textcolor{black}{the} implementation complexity issues of \ac{MLPH} and greedy heuristics. The computation required to solve \ac{MLPH} is too high to cater by low power \textcolor{black}{wireless nodes}. Thus, similar to \cite{Srivastava-Koksal-twcom-15}, we use the memory resource of \textcolor{black}{the} sensor nodes rather than the computational complexity. A look-up table $\mathbf{T}$, which has been pre-computed and stored in the nodes memory, is used to find the adequate transmit power. It contains the actions for different probabilities of \ac{EH}, transmitter side battery, channel, acknowledgement and retransmit index states. The node, at every time slot, updates the channel belief state $\varpi(G_{t})$ and looks up the transmit power $a_{t}$, corresponding to this value.

The memory requirement for storing the look-up table $\mathbf{T}$ depends \textcolor{black}{on} the total number of the system space states $|\mathcal{S}|$ and the number of actions $|\mathcal{U}|$. \textcolor{black}{The} look-up table is stored for different values of the probabilities of \ac{EH}. If \textcolor{black}{each \ac{EH} probability value} is divided into $\kappa$ levels, then the total memory requirement is $\kappa\times|\mathcal{U}|\times|\mathcal{S}|^{2}$ bits. Additionally, $10^{|\mathcal{S}|}$ bits of memory are required to store the belief vector of size $|\mathcal{S}|$, and each element is quantized into 10 levels. On the other hand, the greedy algorithm requires neither computation nor memory resource of \textcolor{black}{the sensors}. It only computes the immediate cost as a function of the current state of the system including the belief state of the channel. This computation has very low complexity when compared to \textcolor{black}{computing the} expected future costs.

\section{\noindent Packet Drop Probability Analysis }
In this section, the queuing process induced by the adaptive retransmission scheme is analyzed for the link between two sensors nodes. In particular, \ac{PDP} is derived by leveraging tools from the queuing theory. The \ac{PDP} is the probability that the transmitted packet \textcolor{black}{has been} dropped due to repeatedly decoding failure or not decoded due to \textcolor{black}{the} lack of energy at the receiver \textcolor{black}{over} $K$ retransmission attempts. In this section, we consider that the channel state remains constant for the duration of one frame transmission and changes to a new state with some transition probability at the start of the new frame. \textcolor{black}{The} \ac{EH} and consumption models are defined in Section II.

In order to make the \ac{PDP} analysis tractable, we consider \textcolor{black}{equal and fixed power policy, where the energy required to transmit a full packet is fixed to $\beta E^{\mathrm{min}}_{\mathrm{Tx}}$. For example, for the case when $\beta = 4$, if the transmitter battery has energy sufficient to transmit only $1/2$ portion of the packet, then the transmit energy is $2E^{\mathrm{min}}_{\mathrm{Tx}}$. On the other hand, if the transmitter battery has less than the $E^{\mathrm{min}}_{\mathrm{Tx}}$, the transmit energy is zero.} Hence, we approximate the system by discrete-time \ac{FSMC}, which has \textcolor{black}{the} state space
$\mathcal{S}=\mathcal{B}_{\mathrm{Tx}}\times\mathcal{B}_{\mathrm{Rx}}\times\mathcal{G}\times\mathcal{Z}\times\mathcal{K}=\{s_{1},s_{2},\ldots,s_{|\mathcal{S}|}\}$.
The state at time $t$ is denoted by $S_{t}=(B^{t}_{\mathrm{Tx}}=i,B^{t}_{\mathrm{Rx}}=j,G_{t}=g,Z_{t}=z,K_{t}=k)$ where $i$, $j$, $g$, $z$, and $k$ are the state values of \textcolor{black}{the} battery at the transmitter and receiver nodes, channel, acknowledgement and retransmission index, respectively.

Let $\bar{P}_{\mathrm{drop}}$ denote the average \ac{PDP}. The packet drop event in a finite battery system is due to either decoding failure or unavailability of energy at the transmitter or at the receiver during $K$ retransmissions.

As depicted in Fig. \ref{fig:TransmissionTimeline}, a frame consists of minimum $1$ to maximum $K$ slots. The acknowledgement state is $z=\mathrm{ACK}$ and the retransmission index state is $k=1$ at the start of the frame. At the end of each slot, $k$ is incremented by 1 if NAKx is received; otherwise, $k=1$ if \ac{ACK} is received. Moreover, after $K$ retransmission attempts, the value of $k$ is reset to 1. If a $\mathrm{NAK}$ is received in the $K$th attempt, then the acknowledge state is reset to $\mathrm{ACK}$ to indicate the start of a next packet transmission.

The \ac{PDP} as a function of $K\geq1$ can be written as
\begin{equation}
\bar{P}_{\mathrm{drop}}(K)=\sum_{i,j}\psi(i,j)\mathbb{E}_{g}\big[P_{\mathrm{drop}}(K|i,j,g,z=\mathrm{ACK},k=1)\big],\label{eq:avg_PDP}
\end{equation}
where $\psi(i,j)$ is the stationary probability that \textcolor{black}{the} transmitter and receiver nodes have energy $iE^{\mathrm{min}}_{\mathrm{Tx}}$ and $jE^{\mathrm{min}}_{\mathrm{Rx}}$, respectively, at the start of the frame. $P_{\mathrm{drop}}(K|i,j,g,z,k)$ is the \ac{PDP} conditioned on the channel gain being in state $G_{t}=g$, transmitter \ac{BSI} $iE^{\mathrm{min}}_{\mathrm{Tx}}$, receiver \ac{BSI} $jE^{\mathrm{min}}_{\mathrm{Rx}}$, acknowledgement state $z$ and retransmission index $k$ at the beginning of the frame. It is given by
\begin{equation}
P_{\mathrm{drop}}(K|i,j,g,\mathrm{ACK},1)=1-P_{\text{\ensuremath{\mathrm{suc}}}},
\end{equation}
where $P_{\mathrm{suc}}$ is the probability that the packet is successfully decoded within $K$ attempts. Thus, $P_{\mathrm{suc}}$ is the sum of all possible events contributing to successful packet transmission. It is given by $P_{\mathrm{suc}}=\sum_{k=1}^{K}P_{\mathrm{suc},k}$, where $P_{\mathrm{suc},k}$ is the probability of success at the $k$th retransmission index. Accounting for \ac{EH} events at the transmitter and receiver EHNs, $P_{\mathrm{suc},k}$ can be \textcolor{black}{upper bounded} as:
\begin{IEEEeqnarray*}{llr}
P_{\mathrm{suc},k}&\leq\Bigg\{1-\bigg[\textcolor{black}{\rho_{\text{Tx}}\rho_{\text{Rx}}}P_{\mathrm{err}}(a_{k})+\textcolor{black}{(1-\rho_{\text{Tx}})\rho_{\text{Rx}}}\varphi_{\mathrm{Tx}}P_{\mathrm{err}}(a_{k}) 
+\textcolor{black}{\rho_{\text{Tx}}(1-\rho_{\text{Rx}})}\Big(P_{\mathrm{err}}(a_{k})(\varphi_{\mathrm{Rx}}+\varphi_{\mathrm{dec}})+\sum_{\mathrm{x}=0}^{\beta}\varphi{}_{\mathrm{Rx,x}}\Big)\\
&+\textcolor{black}{(1-\rho_{\text{Tx}})(1-\rho_{\text{Rx}})}\Big(\varphi_{\mathrm{Tx}}P_{\mathrm{err}}(a_{k})(\varphi_{\mathrm{Rx}}+\varphi_{\mathrm{dec}})
+\sum_{\mathrm{x}=0}^{\beta}\varphi{}_{\mathrm{Rx,x}}\Big)\bigg]\Bigg\}
\times(1-P_{\mathrm{suc},k-1}),\IEEEyesnumber \label{eq:PDP_expression-1}
\end{IEEEeqnarray*}
where $\varphi_{\mathrm{Tx}}=\mathrm{\boldsymbol{1}}_{(a_{k}E^{\mathrm{min}}_{\mathrm{Tx}}\geq E_{\mathrm{Tx}})}$,
$\varphi_{\mathrm{Rx}}=\mathrm{\boldsymbol{1}}_{(jE^{\mathrm{min}}_{\mathrm{Rx}}\geq E_{\mathrm{Rx}})}$,
$\varphi_{\mathrm{dec}}=\mathrm{\boldsymbol{1}}_{(jE^{\mathrm{min}}_{\mathrm{Rx}}\geq E_{\mathrm{dec}})}$
and $\varphi_{\mathrm{Rx,x}}=\boldsymbol{1}_{(\frac{\mathrm{x}+1}{\beta}E_{\mathrm{C,Rx}}>jE^{\mathrm{min}}_{\mathrm{Rx}}\geq\frac{\mathrm{x}}{\beta}E_{\mathrm{C,Rx}})}$. $P_{\mathrm{err}}(a_{k})=P_{\mathrm{err}}(g,a_{k})$, where $a_{k}$ denotes the value of action taken at retransmission time index $k$. Hereafter, the dependency of $g$ is removed from $P_{\mathrm{err}}(g,a_{k})$ since the channel state is assumed fixed during the packet transmission.

The stationary probability distribution $\mathbf{\boldsymbol{\psi}}=[\psi(0,0),\cdots,\psi(i,j),\cdots,\psi(B^{\mathrm{max}}_{\mathrm{Tx}},B^{\mathrm{max}}_{\mathrm{Rx}})]$ can be computed by solving $\mathbf{\boldsymbol{\psi}}=\mathbf{\boldsymbol{\psi}}\mathbf{\boldsymbol{\Psi}}_{g}$,
where $\boldsymbol{\Psi}_{g}$ is the transition probability matrix \textcolor{black}{ whose elements} are given as $\mathbb{E}_{g}\big[\mathrm{Pr}(B^{t+1}_{\mathrm{Tx}}=q,B^{t+1}_{\mathrm{Rx}}=r|B^{t}_{\mathrm{Tx}}=i,B^{t}_{\mathrm{Rx}}=j,g)\big]$, under constraint $\sum_{(i,j)}\psi_{g}(i,j)=1$. Moreover,
\begin{IEEEeqnarray*}{lCr}
\mathbb{E}_{g}\big[\mathrm{Pr}(B^{t+1}_{\mathrm{Tx}}=q,B^{t+1}_{\mathrm{Rx}}=r|B^{t}_{\mathrm{Tx}}=i,B^{t}_{\mathrm{Rx}}=j,g)\big]=\\
\sum_{l=1}^{G}\omega_{o}(g_{l})\mathrm{Pr}(B^{t+1}_{\mathrm{Tx}}=q,B^{t+1}_{\mathrm{Rx}}=r|B^{t}_{\mathrm{Tx}}=i,B^{t}_{\mathrm{Rx}}=j),\quad \IEEEyesnumber \label{eq:channel_expectation}
\end{IEEEeqnarray*}
where the left hand side term is the expected probability that \ac{BSI} of the transmitter and receiver is $q$ and $r$ conditioned on previous \ac{BSI} of $i$ and $j$, respectively. Furthermore, the right hand side term of (\ref{eq:channel_expectation}) can be given as
\begin{IEEEeqnarray*}{llr}
\mathrm{Pr}(B_{t+1}^{\mathrm{Tx}}=q,B_{t+1}^{\mathrm{Rx}}=r|B_{t}^{\mathrm{Tx}}=i,B_{t}^{\mathrm{Rx}}=j)=
\sum_{w}^{|\mathcal{Z}|}\sum_{y=1}^{K}\mathrm{Pr}(q,r,w,y|i,j,z=\mathrm{ACK},k=1).\IEEEyesnumber\label{eq:Trans_Prob_Battery_States}
\end{IEEEeqnarray*}

We use the transition probability matrix $\boldsymbol{\Xi}$ to evaluate $\mathrm{Pr}(B^{t+1}_{\mathrm{Tx}}=q,B^{t+1}_{\mathrm{Rx}}=r|B^{t}_{\mathrm{Tx}}=i,B^{t}_{\mathrm{Rx}}=j,g)$. The elements of matrix $\boldsymbol{\Xi}$ represent the transition probability of going from state $(i,j,z,k)$ to another state $(q,r,w,y)$\textcolor{black}{, which is denoted} by $\mathbf{\boldsymbol{\Xi}}_{i,j,z,k}^{q,r,w,y}$ with fixed $g$. We have identified the following four cases to calculate these elements:

\textit{Case i)} For $z\in\{\mathrm{ACK/NAKx}\},k=1,\ldots,K,$
and \textcolor{black}{both transmitter and receiver are harvesting energy, then 
$\mathbf{\boldsymbol{\Xi}}_{i,j,z,k}^{q,r,w,y}=\textcolor{black}{\rho_{\text{Tx}}\rho_{\text{Rx}}}w_{11}$,}
\textcolor{black}{where $w_{11}$ is described in Table~\ref{tab:CASE-1}.} In this case, the receiver does not feedback NAKx messages and the transmitter resends the full packet since both nodes are harvesting.

\textit{Case ii)} For $z\in\{\mathrm{ACK/NAKx}\}$, and $k=1,\ldots,K$, and the transmitter is harvesting energy, \textcolor{black}{while} the receiver is not, then \textcolor{black}{$
\mathbf{\boldsymbol{\Xi}}_{i,j,z,k}^{q,r,w,y}=\textcolor{black}{\rho_{\text{Tx}}(1-\rho_{\text{Rx}})}w_{10}$}, 
\textcolor{black}{where $w_{10}$ is described in Table~\ref{tab:CASE-2}.} In this case, the receiver can feedback NAKx messages whenever it does \ac{SS}. In response, the transmitter can \textcolor{black}{send} the appropriate fraction of the packet. 
 
\textit{Case iii)} For $z\in\{\mathrm{ACK/NAKx}\}$ and $k=1,\ldots,K$, and the transmitter is not harvesting energy, \textcolor{black}{while} the receiver is harvesting, then \textcolor{black}{$
\mathbf{\boldsymbol{\Xi}}_{i,j,z,k}^{q,r,w,y}=\textcolor{black}{(1-\rho_{\text{Tx}})\rho_{\text{Rx}}}w_{01}$,}
\textcolor{black}{where $w_{01}$ is described in Table~\ref{tab:CASE-3}.} In this case, the receiver never transmits NAKx messages as it is harvesting the entire time slot. The transmitter node can decide to transmit or not depending upon the availability of minimum energy. However, if the current acknowledgement state value is NAKx and the transmitter decides not to transmit, then the next acknowledgement state remains NAKx.

\textit{Case iv)} Both transmitter and receiver are not harvesting energy, and $k=1,\ldots,K$, \textcolor{black}{then $
\mathbf{\boldsymbol{\Xi}}_{i,j,z,k}^{q,r,w,y}=\textcolor{black}{(1-\rho_{\text{Tx}})(1-\rho_{\text{Rx}})}w_{00}$,} 
\textcolor{black}{where $w_{00}$ is described in Table~\ref{tab:CASE-4}.} In this case, assuming $\beta=4$, if the current system state is $S_{t}=(i,j,\mathrm{NAK2},k)$ such that $\varphi_{\mathrm{Tx}}=1$ and $\varphi_{\mathrm{Rx,1}}=1$, then the system moves to state $S_{t+1}=(q,r,\mathrm{NAK1},k+1)$ with probability \textcolor{black}{$(1-\rho_{\text{Tx}})(1-\rho_{\text{Rx}})$.} In another example, if the current system state is $S_{t}=(i,j,\mathrm{NAK},k)$ such that $\varphi_{\mathrm{Tx}}=1$ and $\varphi_{\mathrm{Rx,1}}=1$, the system moves to \textcolor{black}{the} new state $S_{t+1}=(q,r,\mathrm{NAK1},k+1)$ with probability \textcolor{black}{$(1-\rho_{\text{Tx}})(1-\rho_{\text{Rx}})$. The energy levels at the receiver nodes are defined as $a_{\mathrm{Rx}}=E_{\mathrm{Rx}}/E^{\mathrm{min}}_{\mathrm{Rx}}$, $a_{\mathrm{dec}}=\frac{E_{\mathrm{dec}}}{E^{\mathrm{min}}_{\mathrm{Rx}}}$, and $a_{\mathrm{Rx,x}}=\frac{\mathrm{x}E_{\mathrm{C,Rx}}}{\beta E^{\mathrm{min}}_{\mathrm{Rx}}}$.}

\begin{table}[h]
\textcolor{black}{
\begin{subtable}[t]{1\textwidth}
\renewcommand{\arraystretch}{0.65}
\centering
\caption{Case-I}
\label{tab:CASE-1}
\begin{tabular}{c|l}
$w_{11}$ & Conditions \\
\hline
$P_{\text{err}}(a_k)$ &  
\begin{tabular}{ll}
$w=\mbox{NAK}$, $y = \mod(k,K)+1$ &\\
\hline
$q=\min{\{i+L_{\mathrm{Tx}}-a_{k},B^{\mathrm{max}}_{\mathrm{Tx}}\}}$, $r=\min{\{j+L_{\mathrm{Rx}}-a_{\mathrm{Rx}},B^{\mathrm{max}}_{\mathrm{Rx}}\}}$ &
\end{tabular}\\
\hline
$1-P_{\text{err}}(a_k)$ &  
\begin{tabular}{ll}
$w=\mbox{ACK}$, $y = 1$ &\\
\hline
$q=\min{\{i+L_{\mathrm{Tx}}-a_{k},B^{\mathrm{max}}_{\mathrm{Tx}}\}}$, $r=\min{\{j+L_{\mathrm{Rx}}-a_{\mathrm{Rx}},B^{\mathrm{max}}_{\mathrm{Rx}}\}}$ &\\ 
\end{tabular}\\
\hline 
0 & otherwise
\end{tabular}
\end{subtable}}

\begin{subtable}[t]{1\textwidth}
\centering
\textcolor{black}{
\renewcommand{\arraystretch}{0.65}
\centering
\caption{Case-II}
\label{tab:CASE-2}
\begin{tabular}{c|l}
$w_{10}$ & Conditions  \\
\hline
$P_{\text{err}}(a_k)$ & 
\begin{tabular}{ll}
$w=\mbox{NAK}, y = \mod(k,K)+1$ & \\
\hline
$q=\min\{i+L_{\mathrm{Tx}}-a_{k},B^{\mathrm{max}}_{\mathrm{Tx}}\}$, $r={j-a_{\mathrm{Rx}}}$  & \hfill for $z=\mbox{NAK}$, $\varphi_{\mathrm{Rx}}=1$ \\ 
\hline 
$q=\min\{i+L_{\mathrm{Tx}}-a_{k},B^{\mathrm{max}}_{\mathrm{Tx}}\}$, $r=j-a_{\mathrm{dec}}-a_{\mathrm{Rx,x}}$ & for $z=\mbox{NAKx}$,  $\varphi_{\mathrm{dec}}=1$, $\varphi_{\mathrm{Rx,x}}=1$ \end{tabular}\\
\hline 
$1-P_{\text{err}}(a_k)$ & 
\begin{tabular}{ll}$w=\mbox{ACK}, y = 1$ & \\
\hline
$q=\min\{i+L_{\mathrm{Tx}}-a_{k},B^{\mathrm{max}}_{\mathrm{Tx}}\}$, $r={j-a_{\mathrm{Rx}}}$ & \hfill for $z=\mbox{NAK}$, $\varphi_{\mathrm{Rx}}=1$ \\
\hline 
$q=\min\{i+L_{\mathrm{Tx}}-a_{k},B^{\mathrm{max}}_{\mathrm{Tx}}\}$, $r=j-a_{\mathrm{dec}}-a_{\mathrm{Rx,x}}$ & for $z=\mbox{NAKx}$,  $\varphi_{\mathrm{dec}}=1$, $\varphi_{\mathrm{Rx,x}}=1$
 \end{tabular}\\
\hline 
$1$ & 
\begin{tabular}{ll}
$w=\mbox{NAKx}, y = \mod(k,K)+1$ & \\ 
\hline $q=\min\{i+L_{\mathrm{Tx}}-a_{k},B^{\mathrm{max}}_{\mathrm{Tx}}\}$, $r={j-a_{\mathrm{Rx,x}}}$ & for $z=\mbox{NAK,\,NAKx}$, $\varphi_{\mathrm{Rx,x}}=1$ \end{tabular} \\
\hline
0 & otherwise
\end{tabular}}
\end{subtable}

\begin{subtable}[t]{\textwidth}
\centering
\textcolor{black}{
\renewcommand{\arraystretch}{0.65}
\centering
\caption{Case-III}
\label{tab:CASE-3}
\begin{tabular}{c|l}
$w_{01}$ & Conditions  \\
\hline
$P_{\text{err}}(a_k)$ & 
	\begin{tabular}{ll}
	$w=\mbox{NAK},\, y=\mathrm{mod}(k,K)+1$ & \\
	\hline
	$q=i-a_{k}$, $r=\min\{j+L_{\mathrm{Rx}}-a_{\mathrm{Rx}},B^{\mathrm{max}}_{\mathrm{Rx}}\}$  & 			\hfill for $z=\mbox{NAK}$, $\varphi_{\mathrm{Tx}}=1$ \\ 
	\hline 
	$q=i-a_{k}$, $r=\min\{j+L_{\mathrm{Rx}}-a_{\mathrm{Rx}},B^{\mathrm{max}}_{\mathrm{Rx}}\}$ & for 		$z=\mbox{NAKx}$,  $\varphi_{\mathrm{Tx}}=1$ 
	\end{tabular}\\
\hline 
$1-P_{\text{err}}(a_k)$ & 
	\begin{tabular}{ll}$w=\mbox{ACK}, y = 1$ & \\
	\hline
	$q=\min\{i+L_{\mathrm{Tx}}-a_{k},B^{\mathrm{max}}_{\mathrm{Tx}}\}$, $r={j-a_{\mathrm{Rx}}}$ & 			\hfill for $z=\mbox{NAK}$, $\varphi_{\mathrm{Tx}}=1$ \\
	\hline 
	$q=\min\{i+L_{\mathrm{Tx}}-a_{k},B^{\mathrm{max}}_{\mathrm{Tx}}\}$, $r=j-a_{\mathrm{dec}}-				a_{\mathrm{Rx,x}}$ & for $z=\mbox{NAKx}$,  $\varphi_{\mathrm{Tx}}=1$
	 \end{tabular}\\
\hline 
	$1$ & 
	\begin{tabular}{ll}
	$w=\mbox{NAK,\,NAKx}$, $y = \mod(k,K)+1$ & \\ 
	\hline $q=i$, $r=\min\{j+L_{\mathrm{Rx}},B^{\mathrm{max}}_{\mathrm{Rx}}\}$, $\varphi_{\mathrm{Tx}}=0$ \end{tabular} \\
\hline
0 & otherwise
\end{tabular}}
\end{subtable}
\hspace{\fill}

\begin{subtable}[t]{\textwidth}
\centering
\textcolor{black}{
\renewcommand{\arraystretch}{0.65}
\centering
\caption{Case-IV}
\begin{tabular}{c|l}
$w_{00}$ & Conditions  \\
\hline
$P_{\text{err}}(a_k)$ & 
	\begin{tabular}{ll}
	$w=\mbox{NAK},\, y=\mathrm{mod}(k,K)+1$ & \\
	\hline
	$q=i-a_{k}$, $r=j-a_{\mathrm{Rx}}$ & \hfill for $z=\mbox{NAK}$, $\varphi_{\mathrm{Tx}}=\varphi_{\mathrm{Rx}}=1$ \\ 
	\hline 
	$q=i-a_{k}$, $r=j-a_{\mathrm{Rx}}$ & for $z=\mbox{NAKx}$,  $\varphi_{\mathrm{Tx}}=\varphi_{\mathrm{Rx}}=\varphi_{\mathrm{dec}}=1$ 
	\end{tabular}\\
\hline 
$1-P_{\text{err}}(a_k)$ & 
	\begin{tabular}{ll}$w=\mbox{ACK}, y = 1$ & \\
	\hline
	$q=i-a_{k}$, $r={j-a_{\mathrm{Rx}}}$ & \hfill for $z=\mbox{NAK}$, $\varphi_{\mathrm{Tx}}=\varphi_{\mathrm{Tx}}=1$ \\
	\hline 
	$q=i-a_{k}$, $r=j-a_{\mathrm{dec}}$ & \hfill for $z=\mbox{NAKx}$,  $\varphi_{\mathrm{Tx}}=\varphi_{\mathrm{Rx}}=\varphi_{\mathrm{dec}}=1$
	 \end{tabular}\\
\hline 
	$1$ & 
	\begin{tabular}{ll}
	$w=\mbox{NAKx}$, $y = \mod(k,K)+1$ & \\ 
	\hline
	$q=i-a_{k}$, $r=j-a_{\mathrm{Rx,x}}\varphi_{\mathrm{Rx,x}}$, & \hfill for $z=\mbox{NAK}$, $\varphi_{\mathrm{Tx}}=\varphi_{\mathrm{Rx,x}}=1$\\
	\hline
	$q=i-a_{k}$, $r=j-a_{\mathrm{Rx,x}}\varphi_{\mathrm{Rx,x}}$, & \hfill for $z=\mbox{NAKx}$, $\varphi_{\mathrm{Tx}}=\varphi_{\mathrm{Rx,x}}=1$ \\
	\hline
	$q=i$, $r=j$, & \hfill for $z=\mbox{NAKx}$, $\varphi_{\mathrm{Tx}}=0$ 	
	 \end{tabular} \\
\hline
0 & otherwise
\end{tabular}
\label{tab:CASE-4}}
\end{subtable}
\caption{\footnotesize \textcolor{black}{Values of $w_{11}$, $w_{10}$, $w_{01}$, and $w_{00}$.}}
\label{tab:table1}
\end{table}

\section{Numerical Results}
In this section, we evaluate the performance of the adaptive \ac{ACK}/NAKx scheme and power assignment strategy by numerical simulations. Results are compared with the conventional scheme in order to demonstrate the benefits. The conventional retransmission scheme is denoted by \ac{ACK}/\ac{NAK}.

Three metrics are used to evaluate the performance: the average packet transmission time $T_{p}(t)$, which is the average time taken per packet to be successfully delivered; \ac{PDP}, i.e., $P_{\mathrm{drop}}(t)$, the probability of dropping a packet after $K$ retransmission attempts; the spectral efficiency, which is the ratio of the number of successfully transmitted packets to the total number of packets selected from the data buffer to transmit within a fixed transmission time.

The parameters summarized in Table \ref{table_example} are used in numerical simulations unless otherwise mentioned. \textcolor{black}{The probabilities of \ac{EH} for both nodes are assumed to be the same, i.e., $\rho_{\text{Tx}}=\rho_{\text{Rx}}=\rho$.} Note that the slot duration $T_{s}=1$ second, and thus, the power and energy values can be used interchangeably. \textcolor{black}{The} equal power assignment in each time slot is denoted by $P^{\mathrm{e}}_{\mathrm{out}}$. Furthermore, for low values of $\beta$ such as 4, $P_{\mathrm{fb}}$ is assumed negligible in the simulations.

\begin{table}[H]
\renewcommand{\arraystretch}{0.50}
\caption{Simulation Parameters}
\label{table_example}
\centering
\begin{tabular}{c|c}
\hline
\bfseries Parameters & \bfseries Value\\
\hline
\hline
$G$ & $3$\\
\hline
$c$ & $128$ bits\\
\hline
$M$ and $R_c$ & $2$ and $1/2$\\
\hline
$K$ & $4$\\
\hline
$\beta$ & 4 \\
\hline
$T_s$ and $T$ & $1$ s and $150$ s\\
\hline
$\xi$, $\eta$ and $\alpha$ & $3(\sqrt{M}-1)/(\sqrt{M}+1)$, $0.25$ and $1$\\
\hline
$\sigma_n^2$ & $5$ mW \\
\hline
$P_{\mathrm{C,Tx}}$ and $P_{\mathrm{C,Rx}}$ & $0.1$ W \\
\hline
\textcolor{black}{$\rho_{\text{Tx}}$ and $\rho_{\text{Rx}}$} & $[0,1]$ \\
\hline
$P_{\mathrm{out}}$ and $P_{\mathrm{PA}}$& $\{5, 15\}$ mW and $\{10, 30\}$ mW\\
\hline
$E_{\mathrm{Tx}}^{\mathrm{min}}$ and $E_{\mathrm{Rx}}^{\mathrm{min}}$  & $E^t_{\mathrm{Tx}}/ \beta$ and $E^t_{\mathrm{Rx}}/ \beta$\\
\hline
$B^{\mathrm{max}}_{\mathrm{Tx}}$ and $B^{\mathrm{max}}_{\mathrm{Rx}}$  & $6P_{\mathrm{Tx}}$ and $3P_{\mathrm{Tx}}$ \\
\hline
$P_{\mathrm{dec}}$  & $7P_{\mathrm{C,Rx}}$ \\
\hline
$E^{\mathrm{h}}_{\mathrm{Tx}}$ and $E^{\mathrm{h}}_{\mathrm{Rx}}$  & $3P_{\mathrm{Tx}}T_s$ and $1.5P_{\mathrm{Rx}}T_s$ \\
\hline
\end{tabular}
\label{tab:simulation_parameters}
\end{table}
\begin{figure}[h]
\centering
\includegraphics[width=0.85\columnwidth]{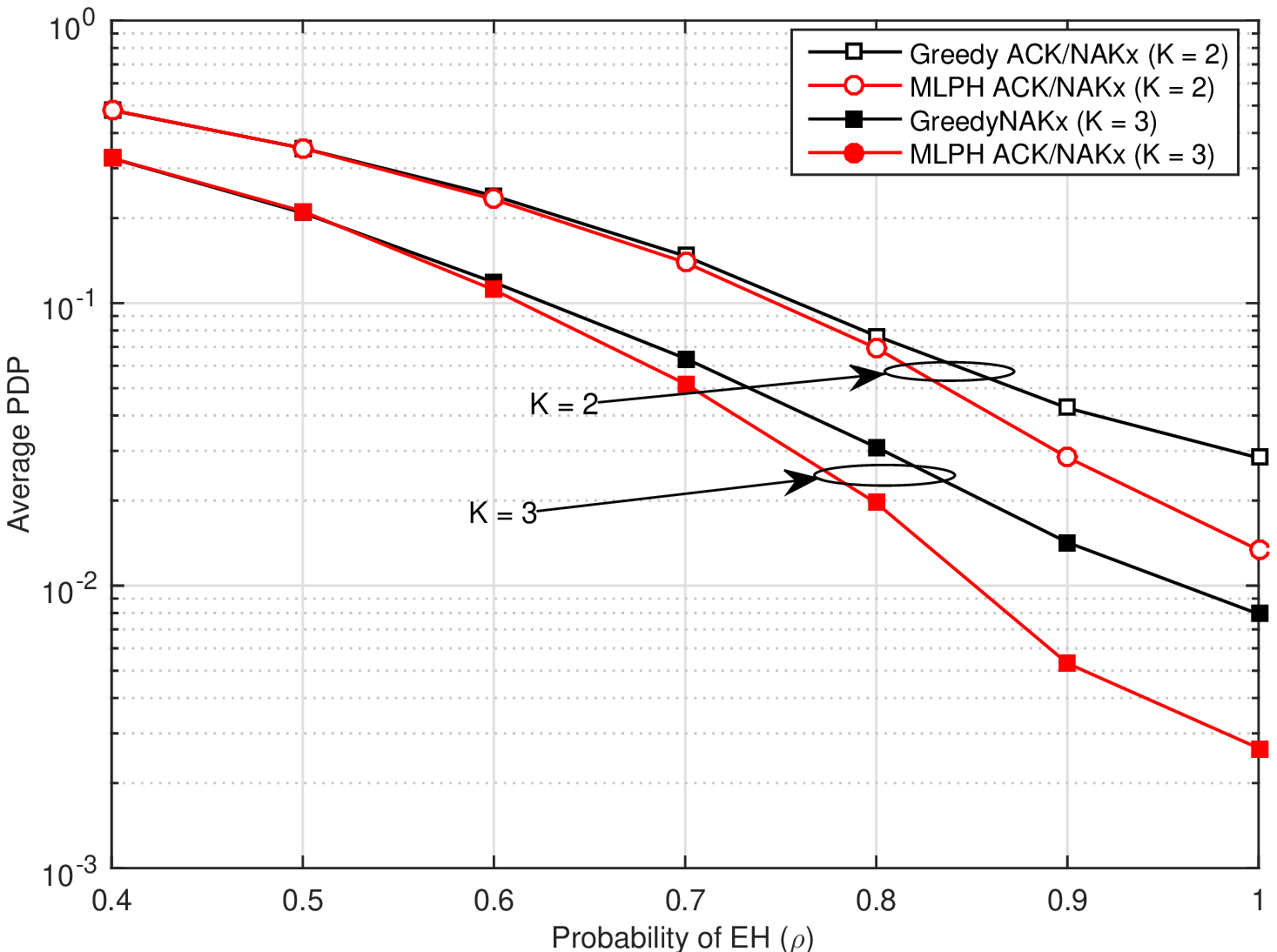}
\protect\caption{Average \acf{PDP} $P_{\mathrm{drop}}(t)$ for \textcolor{black}{$K=\{2,\, 3\}$}.}
\label{MDP_Greedy}
\end{figure}
We first compare the performance of the \ac{MLPH} with the greedy transmit power assignments. Fig. \ref{MDP_Greedy} plots the average \ac{PDP} \textcolor{black}{versus the probability of \ac{EH}. In order to reduce the overall system states, we set the values of $K$ to $2$ and $3$, and $E^{\mathrm{h}}_{\mathrm{Rx}} = 1.2P_{\mathrm{Rx}}T_s$}. The \ac{MLPH} and greedy algorithms have similar performance in lower EH rate regime, whereas the \ac{MLPH} algorithm shows higher gains in higher EH rate regime. \textcolor{black}{As expected, the performances improve with higher number of retransmission attempts, i.e., $K = 3$.} However, when the state size increases, the \ac{MLPH} becomes impractical and \textcolor{black}{the} greedy power assignment strategy becomes a natural choice. Thus, the following numerical examples only consider the greedy approach.

\begin{figure}[h]
\centering
\includegraphics[width=0.85\columnwidth]{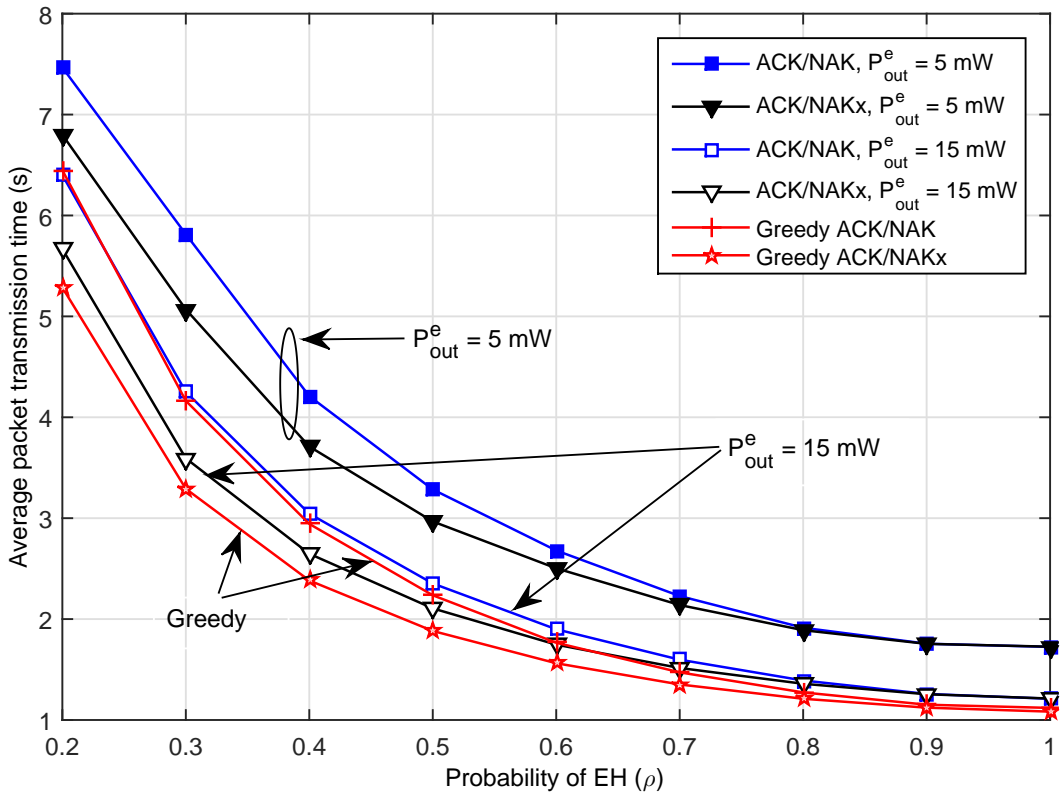}
\caption{Average packet transmission time for $K=4$.}
\label{fig:Time_Greedy_Eq_10_30mW}
\end{figure}
In Fig. \ref{fig:Time_Greedy_Eq_10_30mW}, the average packet transmission time $T_{p}(t)$ is \textcolor{black}{shown versus the probability of \ac{EH}, and} the performance of the \ac{ACK}/NAKx and \ac{ACK}/\ac{NAK} in the greedy and equal power assignment settings\textcolor{black}{, respectively, is compared}. We set $P_{\text{out}}^{\text{e}}=5$ and $15$ mW for equal power assignments. We can observe that the \ac{ACK}/NAKx scheme has the lowest average packet transmission time compared to \textcolor{black}{that} of \ac{ACK}/\ac{NAK} in both greedy and equal power assignment settings. Furthermore, the performance of the \ac{ACK}/NAKx scheme, which employs the equal power assignment of $15$ mW is better than the conventional scheme, which employs the greedy power assignment in the low \ac{EH} regime. As expected, for low harvesting rates, all schemes have longer transmission time. Additionally, the \ac{ACK}/\ac{NAK} scheme \textcolor{black}{exhibits} equal average transmission times for both greedy assignment and equal power assignment with $P_{\text{out}}^{\text{e}}=15$ mW. This result means that \textcolor{black}{a} higher transmit power helps the equal power assignment algorithm in overcoming the channel states that are in deep fade. However, it does not help in using the receiver energy efficiently. Moreover, the performance of \ac{ACK}/NAKx over the \ac{ACK}/\ac{NAK} scheme is significant in lower \ac{EH} rate regime. This is because the receiver node, in the latter scheme, \textcolor{black}{processes the received packet by sampling followed by decoding it}. Due to the lack of energy, the signal processing operation stops, which results in packet drop and loss of energy. Therefore, the receiver has to wait longer to get enough energy to sample and decode the packet in \textcolor{black}{a} single time slot.

\begin{figure}[h]
\centering
\includegraphics[width=0.85\columnwidth]{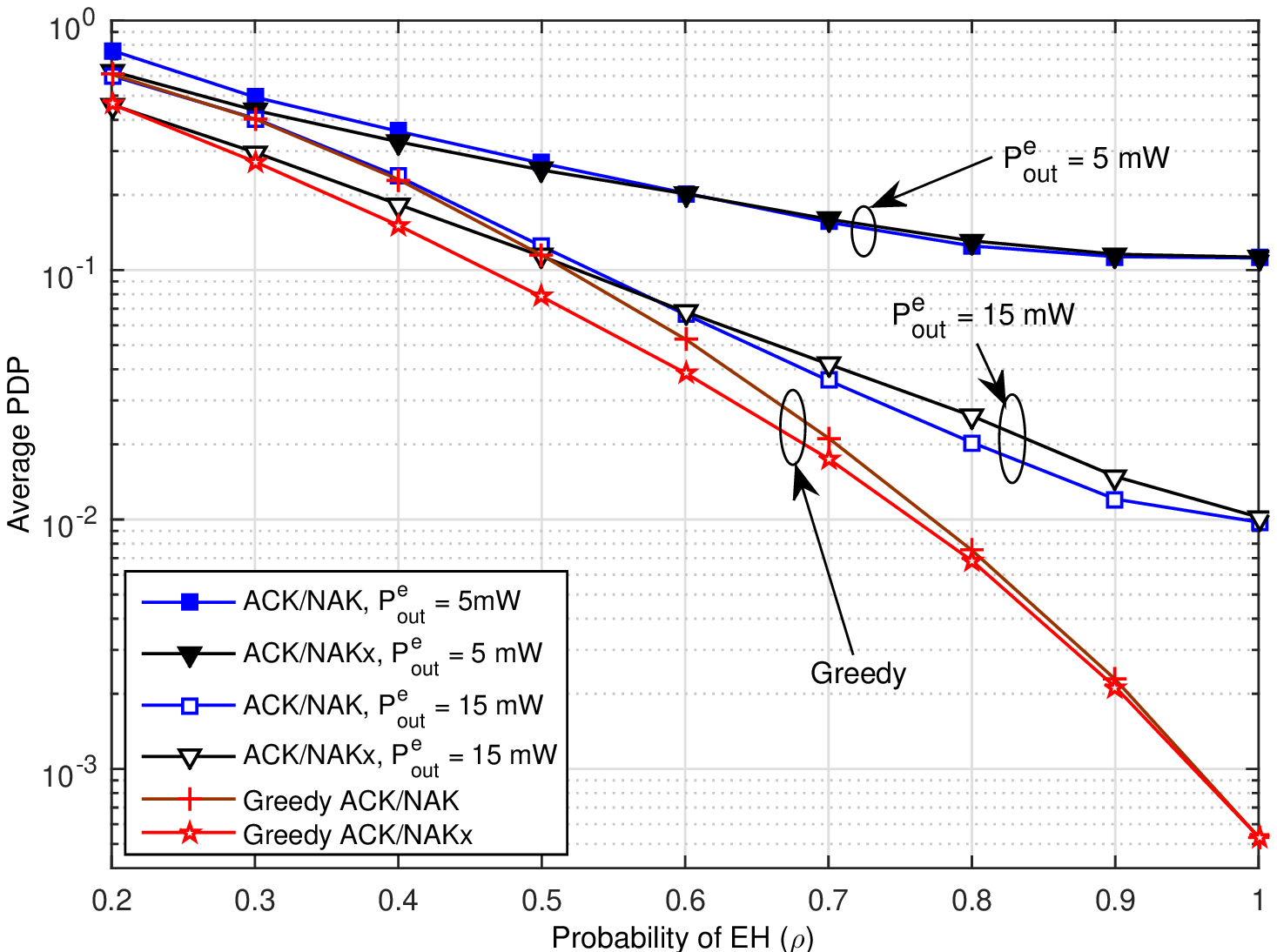}
\protect\caption{Average \acf{PDP} $P_{\mathrm{drop}}(t)$ for $K=4$. }
\label{fig:PDP_Greedy_Eq_10_30mW}
\end{figure}
In Fig. \ref{fig:PDP_Greedy_Eq_10_30mW}, the \ac{PDP} of the \ac{ACK}/NAKx scheme \textcolor{black}{is compared} to that of the \ac{ACK}/\ac{NAK} scheme, \textcolor{black}{when employing the greedy and equal power assignments strategies, respectively.} The simulation parameters are the same as those used for Fig. \ref{fig:Time_Greedy_Eq_10_30mW}. One can see that all retransmission schemes experience high \ac{PDP} in low harvesting rate regime. However, the proposed scheme \textcolor{black}{exhibits} performance gain particularly in low \ac{EH} regime. \textcolor{black}{The} greedy \ac{ACK}/NAKx shows even better gains in all \textcolor{black}{the} harvesting rate regimes.

\begin{figure}[h]
\centering
\includegraphics[width=0.85\columnwidth]{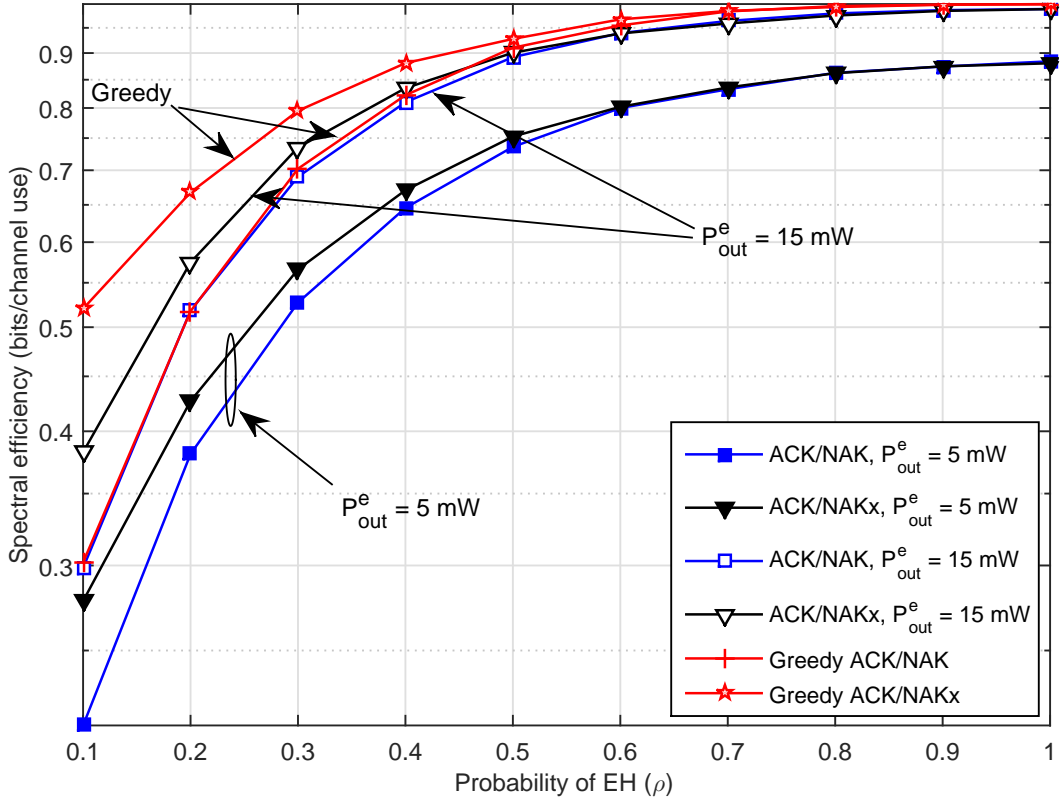}
\caption{Spectral efficiency for a fixed transmission time for $K=4$.}
\label{Spectral_Efficiency_Comparision}
\end{figure}
To get further insight into the performance gain, Fig. \ref{Spectral_Efficiency_Comparision} \textcolor{black}{compares} the spectral efficiency of the \ac{ACK}/NAKx \textcolor{black}{and} the \ac{ACK}/\ac{NAK} schemes for the fixed transmission time $T=150$ s. The \ac{ACK}/NAKx scheme has \textcolor{black}{better} performance over the \ac{ACK}/\ac{NAK} scheme under equal power assignment. Moreover, \textcolor{black}{the} greedy \ac{ACK}/NAKx has better performance over equal power \ac{ACK}/NAKx and \ac{ACK}/\ac{NAK}. Again, as expected, the gains are significant in the low \ac{EH} regime.

\begin{figure}[h]
\centering
\includegraphics[width=0.85\columnwidth]{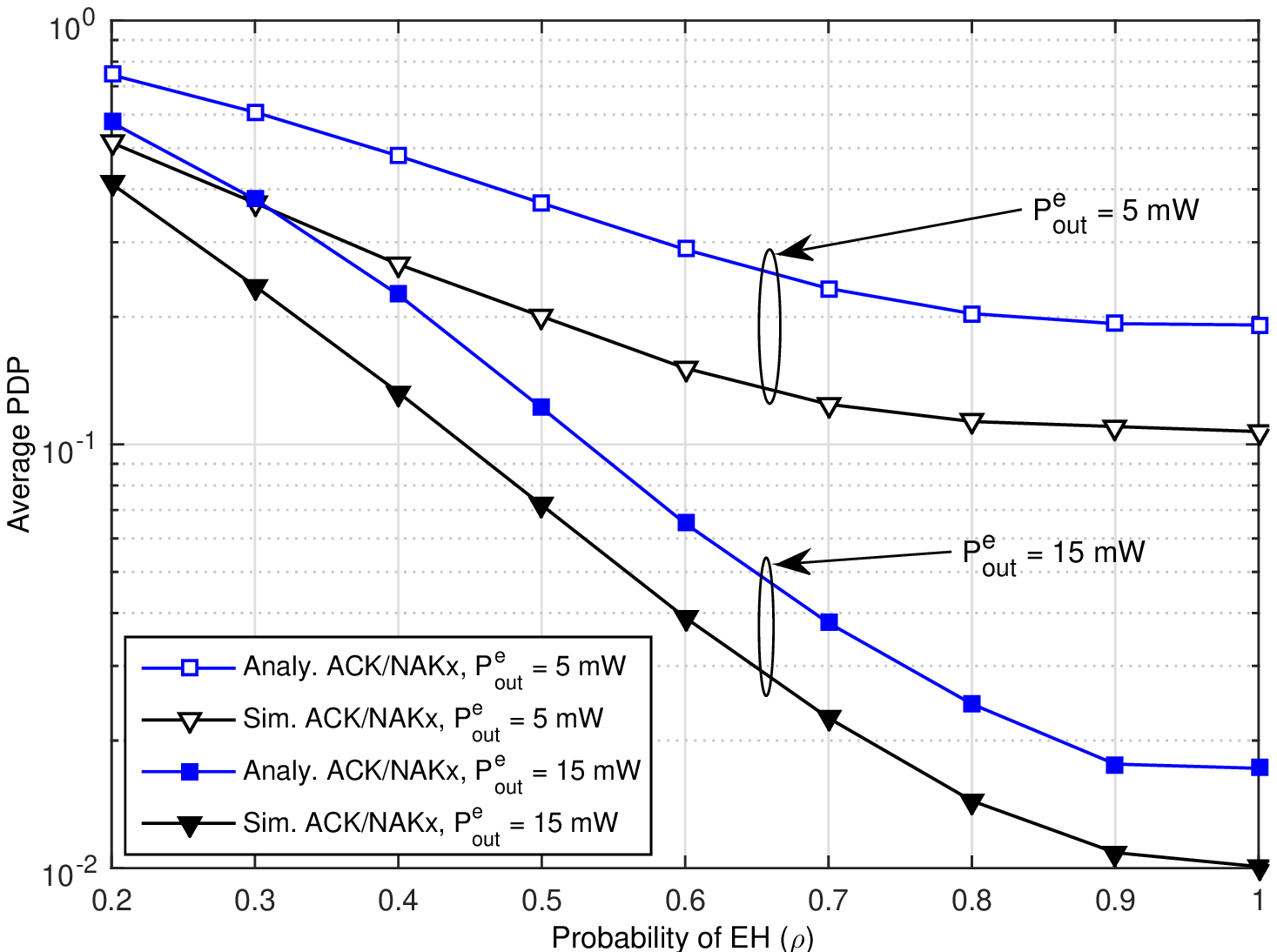}
\caption{Comparison of analytical (\ref{eq:avg_PDP}) and simulated \acf{PDP}.}
\label{fig:Analy_Sim_Comparision}
\end{figure}
In Fig. \ref{fig:Analy_Sim_Comparision}, the simulated \textcolor{black}{and} analytical \textcolor{black}{\ac{PDP} calculated} using (\ref{eq:avg_PDP}) \textcolor{black}{are compared}. In order to reduce the number of states of discrete-time \ac{FSMC}, we set new $E^{\mathrm{h}}_{\mathrm{Tx}}=2P_{\mathrm{Tx}}T_s$,  $B^{\mathrm{max}}_{\mathrm{Tx}}=3P_{\mathrm{Tx}}T_s$, $E^{\mathrm{h}}_{\mathrm{Rx}}=1.2P_{\mathrm{Rx}}T_s$, $B^{\mathrm{max}}_{\mathrm{Rx}}=2P_{\mathrm{Rx}}T_s$, and $P_{\mathrm{dec}}=5P_{\mathrm{C,Rx}}$. \textcolor{black}{It can be seen} that the analytical and simulated curves for $P_{\mathrm{out}}^{\mathrm{e}}\in{\{5,15\}}$ mW have a similar behavior; however, loose. This is because the exact tight bound for \textcolor{black}{the \ac{PDP}} of any modulation with convolution coding is not known for low \ac{SNR}. Consequently, we use the upper bound on the \ac{PEP} (\ref{eq:PacketErrorProb}), which is loose in low and tight in high \ac{SNR} regime, respectively, \textcolor{black}{in} deriving the \ac{PDP} expression (\ref{eq:avg_PDP}). 

\begin{figure}[h]
\centering
\includegraphics[width=0.85\columnwidth]{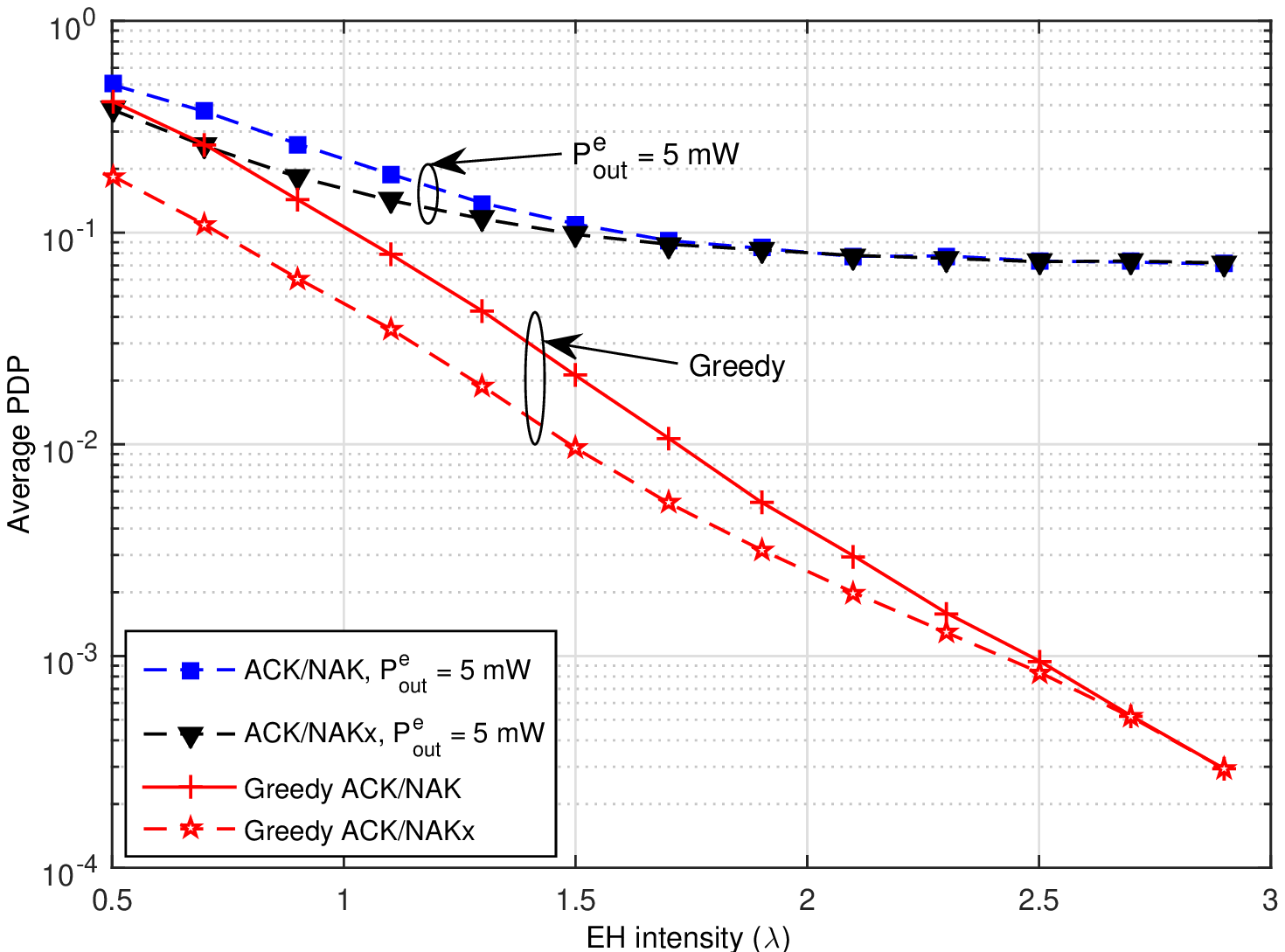}
\caption{Average \acf{PDP} $P_{\mathrm{drop}}(t)$ for $K=4$.}
\label{fig:PDP_Comparision_Poisson}
\end{figure}

For further exposition, Fig. \ref{fig:PDP_Comparision_Poisson} \textcolor{black}{compares} the performance of the \ac{ACK}/NAKx \textcolor{black}{and} \ac{ACK}/\ac{NAK} under a stochastic \ac{EH} setup modeled by the compound Poisson process \cite{Ozel-Tutuncuoglu-Yang-Ulukus-Yener-jsac-11, Xu-Zhang-jsac-14}. The compound Poisson process closely models the \ac{EH} due to the solar power \cite{Bai-Nossek-spawc-13,lee-zhi-mingding-tan-wcnc-2011}. In this model, the energy arrivals follow a Poisson distribution with intensity $\lambda$, i.e., inter-arrival time is exponentially distributed with mean $1/\lambda$. The energy amount in each arrival, i.e., $E_{\mathrm{\{Tx,Rx\}}}^{\mathrm{h}}$ is \ac{i.i.d.}, with mean $\bar{E}_{\mathrm{\{Tx,Rx\}}}$. The number of arrivals in one time slot follows a Poisson distribution with mean $\lambda T_s$. \textcolor{black}{The} simulated \ac{PDP} of the \ac{ACK}/NAKx \textcolor{black}{and} \ac{ACK}/\ac{NAK} \textcolor{black}{are compared} for the greedy and equal power $P_{\mathrm{out}}^{\mathrm{e}} = 5$ mW assignments strategy. The simulation parameters are \textcolor{black}{the} same as the ones in Fig. \ref{fig:Time_Greedy_Eq_10_30mW}, except \textcolor{black}{the} harvesting model with $\bar{E}_{\mathrm{Tx}} = 3P_{\mathrm{Tx}T_s}/2$ and $\bar{E}_{\mathrm{Rx}} = 1.5P_{\mathrm{Rx}T_s}/2$. The behavior of the schemes is \textcolor{black}{the} same as in the case with the Bernoulli EH arrival process.

\textcolor{black}{It is worth mentioning} that the overall gain of the \ac{NAK}/NAKx with low complexity greedy power assignment method is approximately $10-15\%$ when compared to \ac{ACK}/\ac{NAK}. While this gain may appear low, it is significant in \ac{WSN} where \textcolor{black}{multiple} sensors are in operation.

\section{Conclusions \textcolor{black}{and Future Directions}}

\textcolor{black}{An} \ac{ARQ} based adaptive retransmission scheme between a pair of \ac{EH} wireless sensor nodes is investigated. \textcolor{black}{In a conventional scheme,} the receiver may suspend the sampling and decoding operations due to insufficient energy, and hence, suffer a loss of both data and harvested energy. To overcome this problem, a selective sampling scheme was introduced, where the receiver selectively samples the received data and stores it. The selection depends on the amount of energy available. The receiver performs the decoding operation when both complete samples of the packet and enough energy are available. Selective sampling information is fed back to the transmitter by resorting to the \textcolor{black}{conventional} \ac{ARQ} scheme. The transmitter uses \textcolor{black}{this information} to re-size the packet length. A POMDP formulation was setup to further optimize the transmit power. A suboptimal greedy power assignment method was developed, which is well suited for low power wireless nodes from the implementation perspective. An analytical upper bound on the \ac{PDP} is derived for the \textcolor{black}{proposed adaptive retransmission} scheme. Simulation results \textcolor{black}{agree} with the analytical solution when \textcolor{black}{the} fixed power assignment policy is used. Numerical results demonstrated that the adaptive retransmission scheme and power assignment strategy \textcolor{black}{provide} better performance over the conventional scheme.

\textcolor{black}{The proposed adaptive retransmission framework paves the way to several other interesting research avenues. The proposed scheme can be investigated with more sophisticated retransmission scheme, i.e., type-II \ac{HARQ}. To conceptualize the performance of the proposed scheme in a practical scenario, one can consider a system setup with multiple transmitters and one receiver, all with \ac{EH}. Problems to be studied under this setting are more challenging due to the involvement of an increased number of random variables associated to each node.}

\bibliographystyle{IEEEtran}

\end{document}